\UseRawInputEncoding
\documentclass[prx,onecolumn,showpacs,preprintnumbers,amsmath,amssymb,notitlepage]{revtex4-1}
\usepackage{graphicx}
\usepackage{ulem}

\begin{document}


\title{  Response of a strongly interacting spin-orbit coupling system to a Zeeman field }
\author{ Fadi Sun and Jinwu Ye   }
\affiliation{
 Institute for Quantum Science and Engineering, Shenzhen 518055, China   \\
 Department of Physics and Astronomy, Mississippi State University, MS, 39762, USA   }

\date{\today }

\begin{abstract}
 A strongly spin-orbital coupled systems could be in a magnetic ordered phase at zero field. However, a Zeeman field could
 drive it into a topological phases or vice versa.
 In this work, starting from general symmetry principle, we construct various effective actions to study
 the response to a longitudinal Zeeman field of
 strongly interacting spinor atoms with a 2 dimensional (2d) anisotropic Rashba type of spin orbital coupling (SOC) in a square lattice.
 We find the interplay between the Zeeman field and the SOC leads to rich and novel classes of quantum commensurate (C-) and
 in-commensurate (IC-) phases, exotic excitations and novel quantum phase transitions (QPT).
 These phases include the collinear gapped Z-x at low Zeeman field, collinear gapped Z-FM at high Zeeman field,
 gapless co-planar canted phase at low SOC and gapless non-coplanar IC-Skyrmion crystal (IC-SkX) at large SOC in intermediate Zeeman fields.
In the C-IC transition from the Z-x to the IC-SkX at a lower critical field with the dynamic exponent $ z=2 $,
we find a new type of dangerously irrelevant operator we name type-II which is irrelevant near the QPT, but
leads to the exotic excitations inside IC-SkX phase, in distinction from the known one we name type-I.
In the C-IC transition from the the Z-FM to the IC-SKX at a upper critical field with $ z=2 $,
we find the IC- leads to an emergent $ U(1)_{ic} $  symmetry and two type-II dangerously irrelevant operators.
 In the C-C transition from the Z-FM to the canted phase with $ z=1 $, we find the SOC leads to a boost to the 3D XY universality class without SOC. It is the boost which leads to the exotic excitations inside canted phase.
 In the C-IC transition from canted to the IC-SkX with $ (z_x=3/2, z_y=3) $,
 we find an order parameter fractionization where one complex order parameter split into two  which is different than, but related to the quantum spin fractionization into a spinon and a $ Z_2 $ flux.
 We derive the relations between the quantum spin and the order parameters of the effective actions
 which is crucial to determine the spin-orbital structures of all these quantum phases.
 We argue that it is the exotic form of the gapless excitations in the canted or IC-SkX phase which leads to un-quantized thermal Hall conductivities even at zero temperature limit. Finite temperature transitions are presented.
 The dynamic spin-spin correlation functions are evaluated.
 In view of recent experimental advances in generating 2d SOC for cold atoms in optical lattices, these new many-body phenomena
 can be realized in the near future cold atom experiments.
 Implications to various SOC materials such as MnSi, Fe$_{0.5}$Co$_{0.5}$Si,
 especially 4d Kitaev materials $\alpha$-RuCl$_3$ in a Zeeman field are given.
\end{abstract}

\maketitle

\section{Introduction}


During the last decade, the investigation and control of spin-orbital coupling (SOC) have become  the subjects of
intensive research in both condensed matter and cold atom systems after the discovery of the topological insulators
\cite{kane,zhang}. In the condensed matter side, there are increasing number of
new quantum materials with significant SOC, including several new 4d or 5d transition metal oxides and heterostructures of transition metal systems \cite{SLrev1,SLrev3}. In the cold atom side,
several groups worldwide \cite{soexp,sofermigas,sobecgas} have also successfully generated
a 1D  (SOC) to neutral atoms. However, one of the main limitations to extend 1D SOC to a 2D SOC is the associated heating rates.
Recently, there are some advances \cite{expk40,expk40zeeman,clock,2dsocbec,ben} to overcome this difficulty
in generating 2D Rashba SOC for cold atoms in both continuum and optical lattices and also in a Zeeman field.
In view of these recent experimental advances,
it becomes topical and important to investigate what would be new phenomena due to the interplay
among strong interactions, SOC and a Zeeman field in both cold atoms and condensed matter systems.



In \cite{rh}, we studied interacting spinor bosons at integer fillings loaded in a square optical lattice in the presence of non-Abelian gauge fields. In the strong coupling limit, it leads to the spin $ S=N/2 $ Rotated Ferromagnetic Heisenberg model (RFHM)
( Eq.\ref{rhgeneral} with $ \vec{H}=0 $ )
which is a new class of quantum spin models to describe quantum magnetisms in cold atom systems or some materials with strong SOC.
Along the anisotropic line $ (\alpha=\pi/2, 0 < \beta < \pi/2) $ of the 2d SOC, there is an exact $ U(1)_{soc} $ symmetry.
We identified a new spin-orbital entangled commensurate ground state: the Y-x state.
It supports not only commensurate magnons (C$_0$,C$_{\pi}$),
but also a new gapped elementary excitation: in-commensurate magnon ( IC- ).
The IC- magnons may become the seeds to drive possible new classes of quantum C-IC transitions under various external probes.
In \cite{rhh}, by performing the microscopic calculations,  we explored the dramatic effects of
an external longitudinal Zeeman field $H$ applied to the RFHM Eq.\ref{rhgeneral} along the anisotropic SOC line
$ (\alpha=\pi/2, 0 < \beta < \pi/2) $ which keeps the $ U(1)_{soc} $ symmetry.
We find that the interplay among the strong interactions, SOC and the Zeeman field leads to a
whole new classes of magnetic phenomena in quantum phases ( especially the non-coplanar incommensurate Skyrmion  crystals (IC-SkX)  ),
excitation spectra ( especially inside the IC-SkX ), quantum  phase transitions ( especially the quantum Commensurate to incommensurate (C-IC) transitions ), which may have wide and important applications in both cold atoms and various materials with SOC.
Our main results are summarized in  Fig.1.
In \cite{rhtran}, we studied the response to a transverse field of the RFHM Eq.\ref{rhgeneral} along the anisotropic SOC line
$ (\alpha=\pi/2, 0 < \beta < \pi/2) $. Because the  transverse field explicitly breaks the $ U(1)_{soc} $ symmetry,
so the response is quite different than that in a longitudinal field.
However, the approach used in \cite{rh,rhh,rhtran} is exact symmetry analysis plus microscopic spin wave expansion,
so can not be used to study the nature of all these quantum phase transitions. A complete independent symmetry based phenomenological
effective action is needed to achieve this goal.

 In this work, starting from symmetry principle, we construct various effective actions to study all these quantum phases and
 phase transitions in Fig.1.
 We recover all these quantum phases and their excitations discovered by the microscopic calculations in \cite{rh,rhh,rhtran},
 most importantly, explore the nature of all the quantum phase transitions,
 therefore provide deep insights into the global phase diagram in Fig.1.
 Furthermore, we find a new type of dangerously irrelevant operator:
 it is irrelevant near the QCP, but marginal in the symmetry breaking ground state. So it does not change the ground state, but
 changes its excitation spectrum to an exotic form.
 This is in sharp contrast to the known  dangerously irrelevant operator \cite{scaling,sachdev,aue,NOFQD}:
 it is  irrelevant near the QCP, but relevant in the symmetry breaking ground state. So it change both the ground state and the excitation spectrum. We name the known one  and the new one as Type-I and Type-II dangerously irrelevant operator respectively.
 The Z-x state to the IC-SkX transition at $ h=h_{c1} $ is in the same universality class as the $ z=2 $ SF-Mott transition,
 but there is  a type-II  dangerously irrelevant operator which leads to one exotic Goldstone mode
 inside the IC-SkX phase near $ h_{c1} $.
 However, at the mirror symmetry point, the  Type-II dangerously irrelevant operator is absent, the exotic Goldstone mode recovers to the conventional one.  The FM state to the IC-SkX transition at $ h=h_{c2} $ in the middle range
 $ \beta_1 < \beta < \beta_2=\pi/2- \beta_1 $ of SOC is in the same universality class as a $ z=2 $ two-component SF-Mott
 transition in the Ising limit with a $ U(1)_{soc} \times U(1)_{ic} $ symmetry, the extra $ U(1)_{ic} $ symmetry comes from
 the magnon condensation at two IC- momenta.  There are also  two type-II  dangerously irrelevant operators
 which lead to one  exotic gapless Goldstone mode and one gapped exotic roton mode inside the IC-SkX phase near $ h_{c2} $.
 However, at the mirror symmetry point, the two Type-II dangerously irrelevant operators are absent,
 there is a quartic Umklapp term which breaks the extra $ U(1)_{ic} $ symmetry explicitly, the exotic Goldstone and roton mode
 recover to the conventional ones. The FM state to the canted phase transition at $ h=h_{c2} $ in the left ( or ) right range
 $ 0 < \beta < \beta_1 $ ( or $ \beta_2 < \beta < \pi/2 $ ) of SOC is
 in the same universality class of  $ z=1 $ boosted SF-Mott transition.
 It is the SOC which leads to the boost which, in turn, leads to one  exotic Goldstone mode and one exotic Higgs mode inside the canted phase.
 However, at the $ \beta=0 $ Abelian point which maps to a FM in the presence of a staggered Zeeman field along $ x $- axis,
 the boost is absent, the transition reduces to the $ z=1 $ 3d XY class, the exotic Goldstone and Higgs modes recover to the conventional ones.
 Inside the canted phase, as the SOC increases at a fixed Zeeman field, the transition from the canted phase  to the IC-SkX phase
 is a novel class of quantum Lifshitz transition with the anisotropic dynamic exponent  $ ( z_x=3/2, z_y=3 ) $.
 There is an order parameter fractionization (OPF) from one complex order parameter to TWO from the left ( canted to IC-SkX ),
 or equivalently, an order parameter reduction (OPR) from TWO complex order parameters to one from the right ( IC-SkX to canted  ).
 Finite temperature transitions above all these quantum phases and QPTs are presented.
 We also examine carefully the relations between the quantum spins and the order parameters which involve
 linearly the unitary transformation below $ h_{c1} $ and Bogliubov transformation above $ h_{c2} $.
 We also show that these relations still hold phenomenologically when $ h_{c1} < h < h_{c2} $ inside the IC-SkX phase,
 despite the two transformations are not defined anymore in the range of the Zeeman field.
 We argue that it is the exotic form of the gapless Goldstone mode which leads to un-quantized thermal Hall conductivities even at zero
 temperature limit. While the mirror symmetry at $ \beta=\pi/4 $ dictates the vanishing of the thermal Hall conductivities.
 The dynamic spin-spin correlation functions are evaluated by using these relations.
 Transverse fields which explicitly break the $ U(1)_{soc} $ symmetry are also discussed.
 In view of recent impressive experimental advances in generating 2d SOC for cold atoms in optical lattices, these new many-body phenomena
 can be explored in the current and near future cold atom experiments.
 Some implications to various SOC materials such as MnSi, Fe$_{0.5}$Co$_{0.5}$Si with a strong Dzyaloshinskii-Moriya (DM) interaction in a Zeeman field, especially the recently discovered 4d Kitaev materials $ \alpha-Ru Cl_3 $ in a Zeeman field are discussed.
 Some future perspectives are outlined.

Despite there are many previous works on the Boson-Einstein condensation (BEC) of bosons, there are very little works on
magnon condensation in a quantum magnet which is very much different from the BEC.
There is a previous work \cite{z2} phenomenologically assuming the magnon condensation with a $ U(1)_s $  spin-rotation symmetry
is in the same universality class as a 2d $ z=2 $ zero density SF-Mott transition. This spin $ U(1)_s $ symmetry mimics the charge conservation
symmetry of the bosons.
This assumption is confirmed in appendix F.
Our work here in the longitudinal field also has one $ U(1)_{soc} $ symmetry, however, it is a spin-orbital coupled $ U(1)_{soc} $ symmetry, so very much different than the spin $ U(1)_s $ symmetry. Indeed,
as demonstrated in the main text and summarized above, the magnon condensation with SOC is also dramatically different than that
without SOC \cite{z2}. Of course, the BEC of spinor bosons with SOC \cite{pifluxgold,pifluxqsl,NOFQD,SFnon,SFQAH} is
also dramatically different than that without SOC.

Due to the SOC, the response dramatically depends on the orientation of the magnetic field, in the main text, we focus on the longitudinal
field, in the appendix E, we will discuss the two transverse fields.

\begin{figure}
\includegraphics[width=7cm]{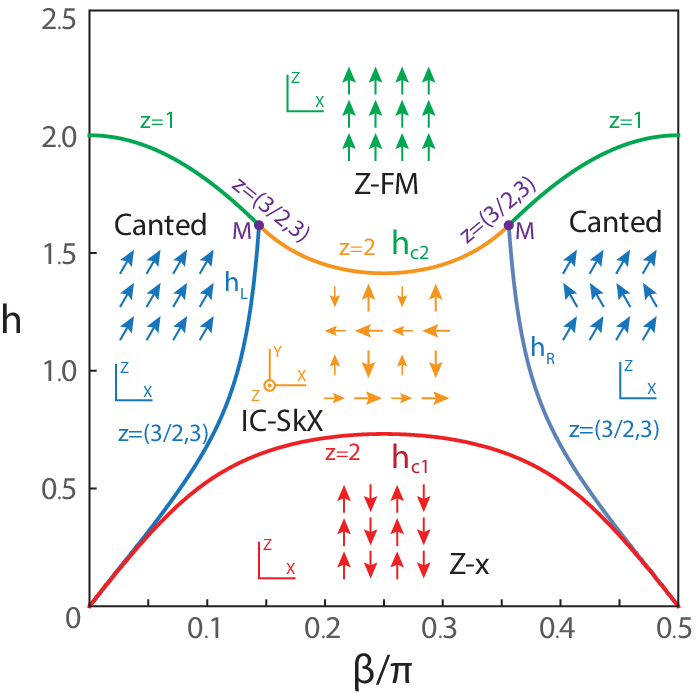}
\hspace{1cm}
\includegraphics[width=8cm]{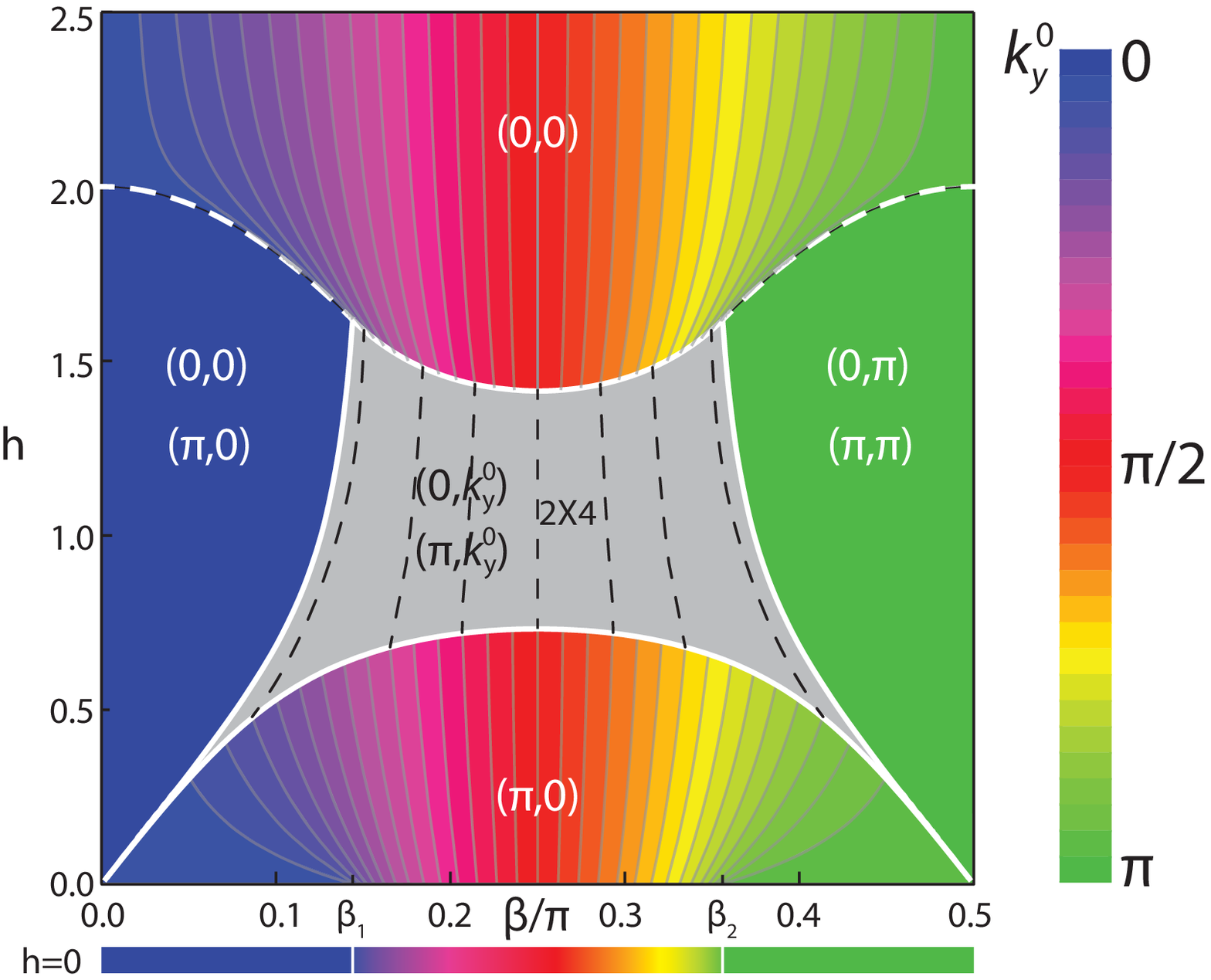}
\caption{ Quantum Phases and phase transitions of RFHM in a longitudinal Zeeman field Eq.\eqref{rhh} achieved by the
combination of the effective actions here and the microscopic  SWE \cite{rhh}
 (a) Below $ h_{c1} $ is the spin-orbital correlated (collinear) Z-x state.
 Above $ h_{c2} $ is the (collinear) Z-FM state. Note the three different pieces of $ h_{c2} $.
 On the left, $ h_L $ is one canted (co-planar) state. On the right, $ h_R $ is another
 canted (co-planar) state. Surrounded by the four commensurate phases is the in-commensurate Skyrmion crystal (non-coplanar) phase (IC-SkX)
 with non-vanishing Skyrmion density.
 There is a multi-critical ( M )  point where the ( collinear ) Z-FM, the ( co-planar ) canted phase and the ( non-co-planar ) IC-SkX phase meet.
 The phases on the left $ \beta < \pi/4 $ are related to the right $ \beta > \pi/4 $ by the Mirror transformation. The center $ \beta = \pi/4 $ respects the Mirror symmetry.  At $\beta=\pi/4$, the IC-SkX reduces to a $2\times4$ commensurate SkX where
 only the spins (with two different lengths) in the $XY$ plane are shown.
 There is one C-C transition from the Z-FM to the canted phase at $ ( h_{c2}, 0 < \beta < \beta_1 ) $ with the dynamic exponent $ z=1 $.
 There are three different kinds of C-IC transitions at $ h_{c1} $, $ ( h_{c2}, \beta_1 < \beta < \beta_2 ) $ and $ h_L $ ( or $ h_R $ )
 from the  Z-x, Z-FM and canted phase to the IC-SkX  with the dynamic exponents $ z=2 $, $ z=2 $ and $ (z_x=3/2, z_y=3 ) $ respectively.
 For the two transverse fields, see appendix E.
(b)  The orbital ordering  wavevectors of  the two collinear, two coplanar and the non-coplanar phases.
The constant contour plot of the minima $ (0, k_y^0 ) $  of the C-IC magnons in the Z-x state at $ h < h_{c1} $ and  Z-FM state
at $ h>h_{c2} $, connected by the orbital ordering wavevectors ( dashed line ) inside the IC-SkX. }
\label{globalphase}
\end{figure}

The spin $ S=N/2 $ Rotated Ferromagnetic Heisenberg model
at a generic SOC parameters $ ( \alpha, \beta) $ in a Zeeman field $ \vec{H} $ along any direction is \cite{rh}:
\begin{eqnarray}
	\mathcal{H}_{RH}  =  -J\sum_i
	[\mathbf{S}_i R(\hat{x},2\alpha)\mathbf{S}_{i+\hat{x}}
	+\mathbf{S}_i R(\hat{y},2\beta)\mathbf{S}_{i+\hat{y}}]
     -  \vec{H} \cdot \sum_i \vec{S}
\label{rhgeneral}
\end{eqnarray}
where  the $ R(\hat{x}, 2 \alpha),  R(\hat{y}, 2 \beta)$ are
two $ SO(3) $ rotation matrices around  the $ \hat{x}, \hat{y} $  spin axis by angle $ 2 \alpha, 2 \beta $
putting along the two  bonds  $ x,y $ respectively, $H$ is the  Zeeman field
which could be induced by the Raman laser in the cold atom set-ups
\cite{expk40,expk40zeeman,clock,2dsocbec,ben}.

Following \cite{rhh}, we focus on studying the phenomena along the line $ (\alpha=\pi/2, 0<\beta<\pi/2 ) $
and in the Zeeman field along the longitudinal $ y $ direction.
After rotating spin $Y$  axis to $Z$ axis by the global rotation  $R(\hat{x},\pi/2)$,
(or equivalently, one can just put $ \beta \sigma_z $  along the $ y $ bonds in the square lattice ),
the Hamiltonian Eqn.\ref{rhgeneral} along the line $ ( \alpha=\pi/2, 0<\beta<\pi/2) $
in the $ H $ field along $ y $ direction can be written as:
\begin{eqnarray}
	\mathcal{H} =  -J\sum_i[\frac{1}{2}(S_i^+S_{i+x}^+ + S_i^-S_{i+x}^-)
	-S_i^zS_{i+x}^z  +  \frac{1}{2}(e^{i2\beta}S_i^+S_{i+y}^-+e^{-i2\beta}S_i^-S_{i+y}^+)
	+S_i^zS_{i+y}^z] -  H \sum_i S_i^z
\label{rhh}
\end{eqnarray}
where the Zeeman field $H$ is along the $ z $ direction after the global rotation.

The symmetry of the Hamiltonian Eq.\ref{rhh} is generated by \cite{symmetrydiffer}
\begin{enumerate}
    \item Translation by one lattice site in $x$ or $y$ direction:
	    $\mathcal{T}_x: S_i\to S_{i+\hat{x}}$
	    and	$\mathcal{T}_y: S_i\to S_{i+\hat{y}}$.
    \item Space reflection with respect to $y$ axis:
	    $\mathcal{I}_y: S_{i}\to S_{\bar{i}}$,
		where $i=(i_x,i_y)$ and $\bar{i}=(-i_x,i_y)$.
    \item Spin reflection symmetry:
	    $\mathcal{P}_z: S_i\to R_z(\pi) S_i$
    \item Spin-orbital reflection:
	    $\mathcal{T}\circ\mathcal{I}_x\circ\mathcal{P}_x$ and
	    $\mathcal{T}\circ\mathcal{I}_x\circ\mathcal{P}_y$.
    \item Spin-orbital coupled $ U(1)_{soc} $ spin-rotation:
	    $\mathcal{R}: S_i\to R_z((-1)^{i_x}\phi)S_{i}$		
    \item* Enlarged mirror symmetry at $ \beta=\pi/4 $:
	    $\mathcal{T}\circ\mathcal{M}$,
	    where $\mathcal{M}: S_i\to R_x(\pi)R_z(i_y\pi) S_i$.
	    It maps Hamiltonian SOC parameter $\beta\to\pi/2-\beta$.
	    $\mathcal{T}\circ\mathcal{M}
	    : (S_i^x,S_i^y,S_i^z)\to(-(-1)^{i_y}S_i^x,(-1)^{i_y}S_i^y,S_i^z)$
\end{enumerate}
    Some of these symmetries are broken in the Z-x, canted and IC-SkX phases, but preserved in the Z-FM state.
    They are quite crucial to construct the corresponding effective actions to be presented in the following.

We will take $ 2SJ $ as the energy unit, so all the physical quantities
such as the Zeeman field $ H $, the magnon dispersion $ \omega_k $ and the gap $ \Delta $  will
be dimensionless after taking their ratios over $ 2SJ $.
We will first focus on the left half of Fig.\ref{globalphase} with $ 0 < \beta < \pi/4 $,
then study the right half using the Mirror transformation $ {\cal M} $.
The mirror center $ \beta=\pi/4 $ respects the Mirror symmetry.


\section{Quantum phase transition at the lower critical field $h_{c1}$}

The spin wave expansion (SWE) in the Z-x state below $ h_{c1} $ was performed in \cite{rhh} and reviewed
in the appendix A.
Dropping the higher branch $\alpha_\mathbf{k}$ in Eq.\ref{betak}, it is the
$\beta_\mathbf{k}$ magnon condensation at $\mathbf{K}_0=(0,k_0)$ which leads to the QPT from the Z-x state to the IC-SkX at $ h_{c1} $ in the whole range of $ 0 < \beta < \pi/2 $. The order parameter  takes the form:
\begin{equation}
\langle \beta_\mathbf{k}\rangle
=\psi\delta_{\mathbf{k},\mathbf{K}_0},~~~~\langle \alpha_\mathbf{k}\rangle=0
\label{drop}
\end{equation}
where $\mathbf{K}_0=(0,k_0)$ and $\psi$ is a complex order parameter.

 One must use the unitary transformation Eq.\eqref{unitaryTrans} to establish the connection between
 the transverse quantum spin and the order parameter:
\begin{align}
    S_{A,i}^+
	=\sqrt{2S}\langle a_i\rangle
	=c\psi e^{ik_0i_y}\quad
    S_{B,j}^-
	=\sqrt{2S}\langle b_j\rangle
	=s\psi e^{ik_0j_y}
\label{k0iyjy}
\end{align}
where $c=c_{\mathbf{K}_0}$ and $s=s_{\mathbf{K}_0}$ are evaluated at $\mathbf{K}_0=(0,k_0)$.
It is easy to see that $ \langle \psi \rangle =0 $ at $ h< h_{c1} $ gives back to the Z-x state.
$ \langle \psi \rangle \neq 0 $ at $ h< h_{c1} $ leads to the IC-SkX state.


The Z-x state spontaneously break the translation along the $x-$ direction by one lattice site
to two lattice site, i.e. $\mathcal{T}_x\to(\mathcal{T}_x)^2$, but still keeps all the other symmetries of the Hamiltonian listed in the introduction. After incorporating this fact, one can study how $\psi$ transform under the symmetries  of the Hamiltonian:
\begin{enumerate}
    \item Translation:
	    $(\mathcal{T}_x)^2: \psi(x,y)\to \psi(x,y)$
	    and $\mathcal{T}_y: \psi(x,y)\to e^{ik_0}\psi(x,y)$;
    \item Space reflection:
	    $\mathcal{I}_y: \psi(x,y)\to \psi(-x,y)$;
    \item Spin reflection:
	    $\mathcal{P}_z: \psi(x,y)\to -\psi(x,y)$;
    \item Spin-orbital reflection:
	    $\mathcal{T}\circ\mathcal{I}_x\circ\mathcal{P}_x:
	    \psi(x,y)\to -\psi^\ast(x,-y)$ and
	    $\mathcal{T}\circ\mathcal{I}_x\circ\mathcal{P}_y:
	    \psi(x,y)\to \psi^\ast(x,-y)$;
   \item Spin-orbital $ U(1)_{soc} $ rotation:
	    $\mathcal{R}: \psi(x,y)\to e^{i\phi_0}\psi(x,y)$;
    \item  Enlarged mirror symmetry at $\beta=\pi/4$:
	    $\mathcal{T}\circ\mathcal{M}:
	    \psi(x,y)\to -\psi^\ast(x,y)$.
\end{enumerate}

Combining the mirror symmetry at  $\beta=\pi/4$ with the spin-orbital reflection leads to the fact that
$\mathcal{I}_x\circ\mathcal{P}_x\circ\mathcal{M}$ maps $\psi(x,y)$ to $\psi(x,-y)$ for $\beta=\pi/4$.
It dictates an odd derivative in $\partial_y$ is  absent at $\beta=\pi/4$, but may appear when away from $\beta=\pi/4$.
The $\mathcal{I}_y$ at a general $\beta$ dictates an odd derivative in $\partial_x$ is always absent.

The above symmetry analysis suggests the following effective action in the continuum limit with the dynamic exponent $z=2$
\begin{align}
	\mathcal{S}_\text{low}
	=	\int d\tau d^2r
	[\psi^\ast\partial_\tau\psi
	+v_x^2|\partial_x\psi|^2
	+v_y^2|\partial_y\psi|^2
	-\mu|\psi|^2
	+U|\psi|^4
	+iV|\psi|^2\psi^*\partial_y\psi +\cdots ]
\label{z2single}
\end{align}

Our microscopic calculation shows that $ \mu=h-h_{c1} $, $U>0 $ and $ V\propto\sin (2k_0)$ which
vanishes at $ \beta=\pi/4 $ dictated by the mirror symmetry.
Due to the factoring of  $ e^{ik_0i_y} $ in Eq.\ref{k0iyjy}, the odd derivative in $ \partial_y $ term first appears in the
interaction $ V $ term.

\subsection{ The spin-orbital order of the IC-SkX state }

At mean field level, we can substitute $\psi =\psi_0=\sqrt{\rho_0}e^{i\phi_0}$
to the effective action Eq.\ref{z2single} and obtain
\begin{align}
    \mathcal{S}_0=-\mu\rho_0+U\rho_0^2
\end{align}
When $\mu<0$ at $ h < h_{c1} $, it is in the Z-x state with $  \langle \psi \rangle =0$.
When $\mu>0$ at $ h > h_{c1} $, it is in the IC-SkX state with $ \langle \psi \rangle =\sqrt{\rho_0}e^{i\phi_0}$
where $\rho_0=\sqrt{\mu/2U}$ and $\phi_0$ is a arbitrary angle due to U(1)$_\text{soc}$ symmetry.

Combining Eq.\ref{k0iyjy} with the constraint $|\mathbf{S}_i|^2=S^2$, one obtain
the spin-orbital order of the IC-SkX phase above $ h_{c1} $:
\begin{eqnarray}
    S_i^+&=&(\sqrt{\rho_0}/2)[c+s+(-1)^{i_x}(c-s)]e^{(-1)^{i_x}i(k_0i_y+\phi_0)}   \nonumber  \\
    S_i^z&=& [\sqrt{S^2-\rho_0c^2}-\sqrt{S^2-\rho_0s^2}
	    +(-1)^{i_x}(\sqrt{S^2-\rho_0c^2}+\sqrt{S^2-\rho_0s^2})]
\label{ordercs}
\end{eqnarray}
where the sign $\pm\sqrt{S^2-|S^+|^2}$ is chosen such
that $S_i^z$ reproduce the Z-x order when $\rho_0\to0$.
It leads to the spin-orbital order in the IC-SkX phase when $h<h_*$ which
is the fixed point in IC-SkX phase where one of the two sublattices $S_i^z=0$.

 One can also calculate
\begin{align}
	\lim_{h\to h_{c1}^-}\frac{|S_A^+|}{|S_B^+|}
	=\lim_{h\to h_{c1}^-}\frac{c}{s}
	=[2-\cos2\beta\cos k_0-\sqrt{(2-\cos2\beta\cos k_0)^2-1}]
\label{hc1ratiocs}
\end{align}
which indeed matches the ratio $|S_A^+|/|S_B^+|$ calculated by SWE from $h^{+}_{c1}$ shown in Eq.\ref{hc1ratio}.

It is important to stress that the quantum spin in Eq.\ref{ordercs} is linearly related to the magnon operator, in contrast to
many other cases where the quantum spin is quadratically represented in terms of spinon operators.
Amazingly, despite the unitary matrix element $  c $ and $ s $ are only well defined inside the Z-x state  below $ h < h_{c1} $.
We can still take them as two phenomenological parameters in Eq.\ref{ordercs} inside the IC-SkX above  $ h > h_{c1} $.
It matches the microscopic SWE calculations in \cite{rhh}.

\subsection{Excitation spectrum: exotic Goldstone mode inside the IC-SkX phase}

When $\mu<0$, $  \langle \psi \rangle =0$ inside the Z-x state,  expanding the effective action upto second order in  $ \psi $ leads to:
\begin{align}
    \mathcal{S}_2
	&=\int d\tau d^2r
	[\psi^\ast\partial_\tau \psi
	+v_x^2|\partial_x \psi|^2
	+v_y^2|\partial_y \psi|^2-\mu| \psi|^2]
\end{align}
which leads to the gapped excitation spectrum
\begin{align}
    \omega_\mathbf{k}
	=-\mu+v_x^2k_x^2+v_y^2k_y^2
\label{Zxex}
\end{align}
  which matches the results achieved by the microscopic SWE calculation in \cite{rhh}.

When $\mu>0$, $  \langle \psi \rangle =  \sqrt{\rho_0}e^{i\phi_0} $ inside the IC-SkX state,
by writing the fluctuations in the polar coordinate
$\psi=\sqrt{\rho_0+\delta\rho}e^{i(\phi_0+\delta\phi)}$, one can expand the action up to the second order in the fluctuations:
\begin{align}
    \mathcal{S}_2
	&=\int d\tau d^2r
	\Big(i\delta\rho\partial_\tau\delta\phi
	+\frac{1}{4\rho_0}[v_x^2(\partial_y\delta\rho)^2+v_y^2(\partial_y\delta\rho)^2]
	+\rho_0[v_x^2(\partial_x\delta\phi)^2+v_y^2(\partial_y\delta\phi)^2]
	+U(\delta\rho)^2
	-V\rho_0\delta\rho\partial_y\delta\phi
	\Big)
\end{align}
 where one can see the odd derivative  in $ \partial_y $ term turns into a quadratic term inside the IC-SkX phase.

Integrating out $\delta\rho$ leads to
\begin{align}
    \mathcal{S}_2
	&=\int d\tau d^2r
	\Big(
	\frac{1}{4U}[(\partial_\tau-i\rho_0V\partial_y)\phi]^2
	+\rho_0[v_x^2(\partial_x\delta\phi)^2+v_y^2(\partial_y\delta\phi)^2]
	\Big)
\end{align}
  where one can see the odd derivative  in $ \partial_y $ term sneaks into $ \partial_\tau $ term inside the IC-SkX phase
  and behaves like a boost term to be discussed in Sec.IV.
  It leads to the exotic Goldstone mode due to the $ U(1)_{soc} $ symmetry breaking:
\begin{align}
    \omega_\mathbf{k}
	=\sqrt{4U\rho_0(v_x^2k_x^2+v_y^2k_y^2)}-V\rho_0 k_y
\label{Goldhc1}
\end{align}
  which recovers the conventional Goldstone mode at  the mirror symmetric point $ \beta=\pi/4 $ where  $ V=0 $.

\subsection{ QPT: Type-II dangerously irrelevant operators away from the mirror symmetric point }

 At the mirror symmetric point $ \beta=\pi/4 $, $ V=0 $, so the
 effective action Eq.\ref{z2single} is in the same universality class as   the $ z=2 $ zero density SF-Mott transition
 where the interaction $ U $ term is marginally irrelevant.
 When away from the mirror symmetric point, the $ V $ term moves in.
 However, simple power counting shows that it is irrelevant near the  $ z=2 $ zero density SF-Mott QCP.
 However, inside the IC-SkX phase, as shown in Eq.\ref{Goldhc1}, it modifies the spectrum of the Goldstone mode
 by an extra linear term, so it is marginal and plays a crucial role inside the phase.
 This is sharp contrast to the well known dangerously irrelevant operator which is irrelevant near the QCP, but
 relevant inside the phase and changes the ground state.  We call this new type of dangerously irrelevant operator Type -II, while the known one as Type-I.  For example, the Type-I appears and leads to the $ N=2 $ XY-AFM phase presented in \cite{NOFQD}.

 So the universality class for the QPT at $h_{c1}$ is nothing but  the $z=2$ 2d SF-Mott transition at the mirror symmetric point,
 plus a Type-II dangerous irrelevant operator away from it.


\section{Quantum phase transition at the upper critical field $h_{c2}$ in the middle range $\beta_1<\beta<\beta_2$. }

The SWE in the FM state above $ h_{c2} $ was also performed in \cite{rhh} and reviewed in appendix B2.
It is the $\alpha_\mathbf{k}$ magnon condensation in Eq.\ref{alphak} which leads to the QPT from the FM state to the IC-SkX at $
 h_{c2} $ in the middle range $\beta_1<\beta<\beta_2$. The order parameter  takes the form:
\begin{equation}
\langle \alpha_\mathbf{k}\rangle =\psi_1\delta_{\mathbf{k},\mathbf{K}_1}
+\psi_2\delta_{\mathbf{k},\mathbf{K}_2}
\end{equation}
where $\mathbf{K}_1=(0,k_0), \mathbf{K}_2=(\pi,k_0)$ and $\psi_1, \psi_2$ are the two complex order parameters.

 One must use the Bogoliubov transformation Eq.\eqref{bogoliubovTrans} to establish the connection between the transverse quantum spin
 and the two complex order parameters:
\begin{align}
	\langle S_i^+\rangle
	\propto
	u[\psi_1+(-1)^{i_x}\psi_2]e^{ik_0i_y}
	+v[\psi_1^\ast-(-1)^{i_x}\psi_2^\ast]e^{-ik_0i_y}
\label{pmk0y}
\end{align}
where $u=u_{\mathbf{K}_1}=u_{\mathbf{K}_2}$
and $v=v_{\mathbf{K}_1}=-v_{\mathbf{K}_2}$.

Because the Z-x state breaks no symmetry of the Hamiltonian, so
one can study how $\psi_1$ and $\psi_2$ transform under the symmetries of the Hamiltonian listed in the Introduction:
\begin{enumerate}
    \item Translation:
	    $\mathcal{T}_x: (\psi_1,\psi_2)(x,y)\to (\psi_1,-\psi_2)(x,y)$
	    and $\mathcal{T}_y: (\psi_1,\psi_2)(x,y)\to (e^{ik_0}\psi_1,e^{ik_0}\psi_2)(x,y)$;
    \item Space reflection:
	    $\mathcal{I}_y:
		(\psi_1,\psi_2)(x,y)
		\to (\psi_1,\psi_2)(-x,y)$;
    \item Spin reflection:
	    $\mathcal{P}_z:
		(\psi_1,\psi_2)(x,y)
		\to (-\psi_1,-\psi_2)(x,y)$;
    \item Spin-orbital reflection:
	    $\mathcal{T}\circ\mathcal{I}_x\circ\mathcal{P}_x:
		(\psi_1,\psi_2)(x,y)
		\to (-\psi_1^\ast,-\psi_2^\ast)(x,-y)$\\ and
	    $\mathcal{T}\circ\mathcal{I}_x\circ\mathcal{P}_y
		: (\psi_1,\psi_2)(x,y)
		\to (\psi_1^\ast,\psi_2^\ast)(x,-y)$;
    \item Spin-orbital $ U(1)_{soc} $ rotation:
	    $\mathcal{R}: (\psi_1,\psi_2)(x,y)
		\to (\psi_1\cos\phi+i\psi_2\sin\phi,
		\psi_2\cos\phi+i\psi_1\sin\phi)(x,y)$;
    \item Enlarged mirror symmetry at $\beta=\pi/4$:
	    $\mathcal{T}\circ\mathcal{M}:
		(\psi_1,\psi_2)(x,y)
		\to(-\psi_1^\ast,-\psi_2^\ast)(x,y)$.
\end{enumerate}
where 
the notation $(\psi_1,\psi_2)(x,y)$ means $(\psi_1(x,y),\psi_2(x,y))$.

Combining the mirror symmetry at  $\beta=\pi/4$ with the spin-orbital reflection leads to the fact that
$\mathcal{I}_x\circ\mathcal{P}_x\circ\mathcal{M}$ maps  $(\psi_1,\psi_2)(x,y)$ to $(\psi_1,\psi_2)(x,-y)$
for $\beta=\pi/4$. It dictates an odd derivative in $\partial_y$ is  absent at $\beta=\pi/4$, but may appear when away from $\beta=\pi/4$.
The $\mathcal{I}_y$ at a general $\beta$ dictates an odd derivative in $\partial_x$ is always absent.
The above symmetry analysis suggests
the following  two-component effective action with the dynamic exponent $z=2$ in the continuum limit,
\begin{align}
    \mathcal{S}_{12}
	=\int d\tau d^2r
	[&\sum_{\alpha=1,2}
	(\psi_\alpha^\ast\partial_\tau\psi_\alpha
	+v_x^2|\partial_x\psi_\alpha|^2
	+v_y^2|\partial_y\psi_\alpha|^2)
	-\mu(|\psi_1|^2+|\psi_2|^2)
	+U(|\psi_1|^2+|\psi_2|^2)^2
	-A(\psi_1\psi_2^\ast+\psi_1^\ast\psi_2)^2\nonumber\\
	&+iV_1(|\psi_1|^2+|\psi_2|^2)
	(\psi_1^\ast\partial_y\psi_1+\psi_2^\ast\partial_y\psi_2)
	+iV_2(\psi_1\psi_2^\ast+\psi_1^\ast\psi_2)
	(\psi_1\partial_y\psi_2^\ast+\psi_1^\ast\partial_y\psi_2)]
\label{12basis}
\end{align}
Our microscopic calculation shows that $ \mu=h_{c2}-h, U=h(u^2+v^2)^2+2(1+h) > A=(4+h) > 0 $.
Furthermore, $ V_1, V_2 \propto\sin (2k_0)$, both of which vanish at $ \beta=\pi/4 $ dictated by the Mirror symmetry.
Due to the factoring out of  $ e^{\pm ik_0i_y} $ in Eq.\ref{pmk0y}, the odd derivative in $\partial_y$ term first appears
in the interaction $ V_1, V_2 $ terms.

In fact, as suggested by Eq.\ref{pmk0y}, the physics may become more transparent in the new basis:
\begin{align}
	\psi_+=(\psi_1+\psi_2)/\sqrt{2},\quad
	\psi_-=(\psi_1-\psi_2)/\sqrt{2}
\end{align}
 where the above effective action becomes \cite{exclude}
\begin{align}
	\mathcal{S}_{\pm}
	=\int d\tau d^2r
	[&\sum_{\alpha=+,-}
	(\psi_\alpha^\ast\partial_\tau\psi_\alpha
	+v_x^2|\partial_x\psi_\alpha|^2
	+v_y^2|\partial_y\psi_\alpha|^2)
	-\mu(|\psi_+|^2+|\psi_-|^2)
	+U(|\psi_+|^2+|\psi_-|^2)^2
	-A(|\psi_+|^2-|\psi_-|^2)^2\nonumber\\
	&+iV_1(|\psi_+|^2+|\psi_-|^2)
	(\psi_+^\ast\partial_y\psi_++\psi_-^\ast\partial_y\psi_-)
	+iV_2(|\psi_+|^2-|\psi_-|^2)
	(\psi_+\partial_y\psi_+^\ast-\psi_-^\ast\partial_y\psi_-)]
\label{pmbasis}
\end{align}
which enjoys a  $ U(1)_{soc} \times U(1)_{ic} $  symmetry when $k_0/\pi$ is an irrational number \cite{irrational}.
The first SOC $ U(1)_{soc} $  maps
$(\psi_+,\psi_-)\to(e^{i\phi_0}\psi_+,e^{-i\phi_0}\psi_-)$, while  the second $ U(1)_{ic} $ is generated by
the whole family of $\mathcal{T}^n_y, n=1,2,3.......$ which maps
$(\psi_+,\psi_-) \to (e^{i k_0 n }\psi_+,e^{i k_0 n}\psi_-)$. Because  $k_0/\pi$ is an irrational number,
so $ \theta_0= k_0 n $ becomes a continuous variable leading to a new emergent $ U(1)_{ic} $ symmetry.

However, if $k_0/\pi=p/q$ with $p$ and $q$ are two coprime positive integers \cite{irrational},
then $(\mathcal{T}_y)^{2q}=1$ and
the action should include an extra Umklapp term:
\begin{align}
	\mathcal{S}_\text{Um}
	&=\int d\tau d^2r\{
	[B_q(\psi_1^2-\psi_2^2)^q+c.c.]
	+[iC_q(\psi_1^2-\psi_2^2)^{q-1}
	(\psi_1\partial_y\psi_2-\psi_2\partial_y\psi_1)+c.c.]+\cdots\}  \\ \nonumber
	&=2^q\int d\tau d^2r\{
	[B_q(\psi_+\psi_-)^{q}+c.c.]
	+[iC_q(\psi_+\psi_-)^{q-1}(\psi_+\partial_y\psi_-)+c.c.]+\cdots\}
\label{umterm}
\end{align}
which breaks explicitly only the $ U(1)_{ic} $, but not the $ U(1)_{soc} $ symmetry. The $B_q,C_q$  maybe complex for $\beta\neq\pi/4$  and $\cdots$ means high order terms with power $2nq$ ($n>1$).

At the mirror symmetric point $ \beta=\pi/4 $,  $k_0=\pi/2$ with $q=2$, then $\mathcal{S}_\text{Um}$ is quartic order in $\psi_{1,2}$ .
So one must consider this $ B_2 $ term at $ \beta=\pi/4 $ where the mirror symmetry dictates $ C_2=0 $ and also the absence of the
two type-II  dangerously irrelevant $ V_1,V_2 $ terms.
\begin{align}
	\mathcal{S}_M
	  =\int d\tau d^2r
	[&\sum_{\alpha=+,-}
	(\psi_\alpha^\ast\partial_\tau\psi_\alpha
	+v_x^2|\partial_x\psi_\alpha|^2
	+v_y^2|\partial_y\psi_\alpha|^2)
	-\mu(|\psi_+|^2+|\psi_-|^2)
	+U(|\psi_+|^2+|\psi_-|^2)^2  \\  \nonumber
	& -A(|\psi_+|^2-|\psi_-|^2)^2+ B_2 (\psi_+\psi_-)^{2}+c.c. ]
\label{pmbasisM}
\end{align}

In the regime $0\leq k_0\leq\pi/2$ in Fig.2a, $q\geq 2$, so $\mathcal{S}_\text{Um}$  becomes higher order
when $\beta<\pi/4$ with $ q>2 $. Then it become highly irrelevant in the renormalization group (RG) sense, so can be dropped \cite{contrastNOFQD}.

\subsection{ The spin-orbital order of the ground state }

The $ \psi_{\pm} $ basis is good for symmetry analysis （ see Sec.C）. However,
the saddle point solution $ (\langle\psi_-\rangle=0, \langle\psi_+\rangle \neq 0 $ or
$ ( \langle\psi_-\rangle \neq 0, \langle\psi_+\rangle=0 ) $ inside the IC-SkX phase,
so it is not convenient to investigate quantum fluctuations in the polar coordinate \cite{singular}. Here, we get back to the $(\psi_1,\psi_2)$ basis.
At mean-field level, we can substitute
$\psi_\alpha\to\sqrt{\rho_\alpha}e^{i\phi_\alpha}, \alpha=1,2$
to the effective action Eq.\ref{12basis}
\begin{align}
    \mathcal{S}_0\propto
	-\mu(\rho_1+\rho_2)
	+U(\rho_1+\rho_2)^2
	-4A\rho_1\rho_2\cos^2(\phi_1-\phi_2)   \\ \nonumber
	=-\mu(\rho_++\rho_-)
	+U(\rho_++\rho_-)^2
	-A(\rho_+-\rho_-)^2
\end{align}

When $\mu=h_{c2}-h <0$, it is in the Z-FM phase with $ \langle \psi_1 \rangle = \langle \psi_2 \rangle=0$.
When $\mu>0$, it is in the IC-SkX phase with $ \langle \psi_1 \rangle= \langle \psi_2 \rangle=\sqrt{\rho_0/2}e^{i\phi_0}$
and $\rho_1=\rho_2=\rho_0/2=\sqrt{\mu/8(U-A)}$.
It is easy to see the symmetry breaking pattern is described by the coset \cite{socsdw}:
\begin{equation}
 U(1)_{soc} \times U(1)_{ic}/[U(1)_{soc} \times U(1)_{ic}]_D
\label{coset}
\end{equation}
where the diagonal ( D ) means $ y \to y+n, \phi_0 \to \phi_0 - n k^0_y $
generated by $ \mathcal{T}^{n}_y \times \mathcal{R}( n k^0_y ) $ for any integer $ n $ \cite{iclead}.
The coset dictates only one Goldstone mode. Note that the IC-SkX phase breaks all other symmetries
of the Hamiltonian  except $\mathcal{I}_x$ and  $ [U(1)_{soc} \times U(1)_{ic}]_D $.

  For the commensurate case $k_0/\pi=p/q$, we may also include the Umklapp contribution:
\begin{align}
    \mathcal{S}_0\propto
	-\mu(\rho_1+\rho_2)
	+U(\rho_1+\rho_2)^2
	-4A\rho_1\rho_2\cos^2(\phi_1-\phi_2)
	+B_q[(\rho_1e^{i2\phi_1}-\rho_2e^{i2\phi_2})^q
	    +c.c.]
\end{align}

When $A\gg|B_q|$, the mean field solution $ \langle \psi_1 \rangle = \langle \psi_2 \rangle=0$ for $\mu<0$
and $ \langle \psi_1 \rangle= \langle \psi_2 \rangle=\sqrt{\rho_0/2}e^{i\phi_0}$ for $\mu>0$ still holds.
This fact can be best seen in the $(\psi_+,\psi_-)$ basis:
\begin{align}
    \mathcal{S}_0\propto
	-\mu(\rho_++\rho_-)
	+(U-A)(\rho_++\rho_-)^2
	+4\rho_+\rho_-\{A+2B_q(4\rho_+\rho_-)^{q/2-1}\cos[q(\phi_++\phi_-)]\}
\end{align}
 where $ \rho_+ + \rho_-=\rho_0 $.

When $A>2|B_q|\rho_0^{q-2} $,
the last term is always non-negative which ensures  $\rho_+\rho_-=0$ in the mean field ground-state.

Combing Eq.\ref{pmk0y}  with the constraint $|\mathbf{S}_i|^2=S^2$, one obtain the spin-orbital order of the IC-SkX phase below $ h_{c2} $
\begin{eqnarray}
    S_i^+&=&\sqrt{\rho_0/2}[u+v+(-1)^{i_x}(u-v)]e^{(-1)^{i_x}i(k_0i_y+\phi_0)} \\ \nonumber
    S_i^z&=&[\sqrt{S^2-2\rho_0u^2}+\sqrt{S^2-2\rho_0u^2}
	+(-1)^{i_x}(\sqrt{S^2-2\rho_0u^2}-\sqrt{S^2-2\rho_0u^2})]/2
\label{orderuv}
\end{eqnarray}
where the sign $\pm\sqrt{S^2-|S^+|^2}$ is chosen such
that $S_i^z$ reproduce the Z-FM when $\rho_0\to0$.
It leads to the spin-orbital order in the IC-SkX phase when $ h_*< H  < h_{c2} $ which
is the fixed point in the IC-SkX phase where one of the two sublattices $S_i^z=0$.

After identifying the  even/odd $i_x$ to be $A$/$B$ sub-lattice, one can also calculate
\begin{equation}
	\lim_{h\to h_{c2}^+}\frac{|S_A^+|}{|S_B^+|}
	=\lim_{h\to h_{c2}^+} \frac{v}{u}
	=\sqrt{\sin^42\beta+\sin^22\beta}
	-\sqrt{\sin^42\beta-\cos^22\beta}
\label{hc2ratiouv}
\end{equation}
which indeed matches the ratio $|S_A^+|/|S_B^+|$
calculated using the SWE from below $ h^{-}_{c2}$ shown in Eq.\ref{hc2ratio}.

It is important to stress that the quantum spin in Eq.\ref{pmk0y}
( or Eq.\ref{orderuv} ) is linearly related to the magnon operator, in contrast to
many other cases where the quantum spin is quadratically represented in terms of spinon operators.
Amazingly, despite the Bogliubov transformation matrix element $  u $ and $ v $ are only well defined above $ h > h_{c2} $.
We can still take the two as two phenomenological parameters in Eq.\ref{pmk0y} ( or Eq.\ref{orderuv} ) inside the IC-SkX below  $ h < h_{c2} $.
It indeed matches the microscopic calculation using SWE in \cite{rhh}.
Note that Eq.\ref{orderuv} takes the identical form as Eq.\ref{ordercs} after replacing the
Bogliubov transformation matrix elements $ u,v$ by the unitary transformation matrix elements $ c,s $.
It is remarkable that one can extend the unitary transformation matrix elements $ c,s $ in the Z-x phase above $ h_{c1} $ and
the Bogliubov transformation matrix elements $ u, v $ in the FM state below $ h_{c2} $ and reach the same spin-orbital
structure of the IC-SkX phase in Eq.\ref{ordercs} and Eq.\ref{orderuv} respectively.

\subsection{Excitation spectrum: exotic gapless Goldstone and gapped roton mode }

When $\mu<0$, it is in the Z-FM state with $ \langle \psi_{\alpha} \rangle = 0, \alpha=1,2 $, expanding the effective action
 upto the second order in $\psi_\alpha$ leads to:
\begin{align}
    \mathcal{S}_2
	=\int d\tau d^2r
	\sum_{\alpha=1,2}
	(\psi_\alpha^\ast\partial_\tau \psi_\alpha
	+v_x^2|\partial_x \psi_\alpha|^2
	+v_y^2|\partial_y \psi_\alpha|^2
	-\mu|\psi_\alpha|^2)
\end{align}
which lead to 2 degenerate gapped modes
\begin{align}
	\omega_{1,2}=-\mu+v_x^2k_x^2+v_y^2k_y^2
\label{gapmiddle}
\end{align}
 which matches the result achieved by SWE in \cite{rhh}.

When $\mu>0$, it is in IC-SkX state with $ \langle  \psi_\alpha  \rangle = \sqrt{\rho_\alpha}e^{i\phi_\alpha}, \alpha=1,2$,
one may write the fluctuations in the polar coordinate as
$\psi_\alpha=\sqrt{\rho_0/2+\delta\rho_\alpha}e^{i(\phi_0+\delta\phi_\alpha)}$
and expand the action upto the second order in the fluctuations.
It turns out to be convenient to introduce
$\delta\rho_\pm=(\delta\rho_1\pm\delta\rho_2)/\sqrt{2}$
and $\delta\phi_\pm=(\delta\phi_1\pm\delta\phi_2)/\sqrt{2}$  where the action becomes
\begin{align}
    \mathcal{S}_2
	&=\int d\tau d^2r\Big(
	i\delta\rho_+\partial_\tau\delta\phi_+
	+\frac{1}{2\rho_0}
	[v_x^2(\partial_x\delta\rho_+)
	+v_y^2(\partial_y\delta\rho_+)]
	+\frac{\rho_0}{2}
	[v_x^2(\partial_x\delta\phi_+)
	+v_y^2(\partial_y\delta\phi_+)]
	 +2(U-A)(\delta\rho_+)^2       \\ \nonumber
	&+i\delta\rho_-\partial_\tau\delta\phi_-
	+\frac{1}{2\rho_0}
	[v_x^2(\partial_x\delta\rho_-)
	+v_y^2(\partial_y\delta\rho_-)]
	+\frac{\rho_0}{2}
	[v_x^2(\partial_x\delta\phi_-)
	+v_y^2(\partial_y\delta\phi_-)]
	+2A(\delta\rho_-)^2
	+2A\rho_0^2(\delta\phi_-)^2      \\  \nonumber
	 &-V_1\rho_0[4\delta\rho_+\partial_y\delta\phi_+
	+2\delta\rho_-\partial_y\delta\phi_-]
	-V_2\rho_0[4\delta\rho_+\partial_y\delta\phi_+
	-2\delta\rho_-\partial_y\delta\phi_-]\Big)
\end{align}
  which leads to one exotic gapless Goldstone and one exotic gapped roton mode
\begin{eqnarray}
    \omega_{+,\mathbf{k}}
	&=&\sqrt{4\rho_0(U-A)(v_x^2k_x^2+v_y^2k_y^2)}
	-(4V_1+2V_2)\rho_0k_y,      \\  \nonumber
    \omega_{-,\mathbf{k}}
	&=&\sqrt{16\rho_0^2A^2+8\rho_0A(v_x^2k_x^2+v_y^2k_y^2)}
	-(4V_1-2V_2)\rho_0k_y
\label{GoldRoton}
\end{eqnarray}
  where the Goldstone mode achieved from below $ h_{c2} $ takes the same form as that in Eq.\ref{Goldhc1}
  achieved from above $ h_{c1} $. While the gapped roton mode corresponds to the higher branch $\alpha_\mathbf{k}$
  in Eq.\ref{drop} which is ignored in the effective action Eq.\ref{z2single}.
  This match is a good check on the consistency between
  the effective action from $ h_{c2} $ down and that from $ h_{c1} $ up.

At the mirror symmetric point $\beta=\pi/4$ ( $k_0=\pi/2$ ) which dictates $V_1=V_2=C_2=0$.
Eq.\ref{pmbasisM} in the $ \psi_{1,2} $ representation becomes:
\begin{eqnarray}
    \mathcal{S}_2
	& = &\int d\tau d^2r\Big(
	i\delta\rho_+\partial_\tau\delta\phi_+
	+\frac{1}{2\rho_0}
	[v_x^2(\partial_x\delta\rho_+)
	+v_y^2(\partial_y\delta\rho_+)]
	+\frac{\rho_0}{2}
	[v_x^2(\partial_x\delta\phi_+)
	+v_y^2(\partial_y\delta\phi_+)]
	+2(U-A)(\delta\rho_+)^2   \\  \nonumber
	& + & i\delta\rho_-\partial_\tau\delta\phi_-
	+\frac{1}{2\rho_0}
	[v_x^2(\partial_x\delta\rho_-)
	+v_y^2(\partial_y\delta\rho_-)]
	+\frac{\rho_0}{2}
	[v_x^2(\partial_x\delta\phi_-)
	+v_y^2(\partial_y\delta\phi_-)]
	+2A(\delta\rho_-)^2
	+2A\rho_0^2(\delta\phi_-)^2 \\  \nonumber
	& + & 4B_2\cos4\phi_0[(\delta\rho_-)^2-\rho_0^2(\delta\phi_-)^2]
	-8B_2\sin4\phi_0(\delta\rho_-)(\delta\phi_-)\Big)
\end{eqnarray}
  where one can see the $ B_2 $ term are endowed with a $ \phi_0 $ dependence and
  only affects the gapped roton $ - $ mode, but not the gapless Goldstone $ + $ mode.
  This is expected, because this $ B_2 $ term breaks only the $ U(1)_{ic} $, but not the  $ U(1)_{soc} $ symmetry.

   The excitations can be extracted as:
\begin{align}
    \omega_{+,\mathbf{k}}
	&=\sqrt{4\rho_0(U-A)(v_x^2k_x^2+v_y^2k_y^2)}, \\  \nonumber
    \omega_{-,\mathbf{k}}
	&=\sqrt{16\rho_0^2(A^2-4B_2^2)+8\rho_0A(v_x^2k_x^2+v_y^2k_y^2)}
\end{align}
 which recover to the conventional form and are independent of $ \phi_0 $ as expected.
 It also indicate the Umklapp term at $ \beta=\pi/4 $ does not affect the Goldstone mode, but decrease the roton gap.


\subsection{ QPT: Two Type-II dangerously irrelevant operators away from the mirror symmetric point  }

 When away from the mirror symmetric point, the Umklapp terms drop out, but the $ V_1, V_2 $ term move in.
 The symmetry is enlarged to $ U(1)_{soc} \times U(1)_{ic} $ which is
 spontaneously broken down to $ [ U(1)_{soc} \times U(1)_{ic}]_D $ in the IC-SkX phase leading to
 one Goldstone mode. In fact, there is also a $ Z_2 $ exchange symmetry between $ \psi_1 $ and $ \psi_2 $
 ( or $ \psi_+ $ and $ \psi_- $ ) which is also broken inside the IC-SkX phase.
 The universality class can be best seen in the $ \psi_{\pm} $ basis Eq.\ref{pmbasis}.
 Because it is the Ising limit, so the saddle point solution
 $ \langle\psi_-\rangle=0 $ or $\langle\psi_+\rangle=0 $ still respects $ [ U(1)_{soc} \times U(1)_{ic}]_D $
 generated by $ \mathcal{T}^{n}_y \times \mathcal{R}( n k^0_y ) $. The two different solutions correspond to
 the exchange of A and B sublattices in the IC-SkX phase. As shown above, the Two Type-II dangerously irrelevant
 operators $ V_1, V_2 $ modify both the Goldstone and the roton mode to the exotic form.

 In one appendix of \cite{pifluxqsl}, we studied the SF-Mott transition in a one component boson at integer filling
 subject to a $ \pi $ flux and reached the same effective action as Eq.\ref{pmbasis} upto to the quartic order, also in Ising limit.
 However, there are no dangerously irrelevant operators.
 In \cite{dual1}, we studied the SF to charge density wave (CDW) transition
 one component boson at half filling in a honeycomb lattice with nearest neighbor repulsive interaction.
 We also reached a similar effective action as Eq.\ref{pmbasis}, also in the Ising limit, with $ \psi_{\pm} $ standing for
 the vortex degree of freedoms hopping in a dual triangular lattice which couple to a gapless fluctuating $ U(1) $ gauge field.
 The saddle point solution  $ \langle\psi_-\rangle=0, \langle\psi_+\rangle \neq 0 $ or
 $ \langle\psi_-\rangle \neq 0, \langle\psi_+\rangle=0 $ correspond to the two CDW states which breaks
 the $ U(1) $ gauge symmetry, open a gap through the Higgs mechanism. There are no dangerously irrelevant operators either.

 At the mirror symmetric point $ \beta=\pi/4 $, the $ V_1, V_2 $ term drop out, but the Umklapp term Eq.\ref{umterm} move in Eq.\ref{pmbasisM}.
 It remains in the Ising limit where one of $ \psi_{\pm} $ vanishes. So the Umklapp term will not change the universality class.
 Due to the absence of the two Type-II dangerously irrelevant operators, the Goldstone and roton modes recover to the conventional ones.

\section{ Quantum phase transition at $h_{c2}$ and in the left range $0<\beta<\beta_1$: Order parameter reduction }

The SWE in the FM state above $ h_{c2} $ leads to Eq.\ref{alphak}.
It is the $\alpha_\mathbf{k}$ magnon condensation which leads to the QPT from the FM state to the canted phase at $
 h_{c2} $ in the left range $ 0<\beta<\beta_1 $. In contrast to the middle range presented in the previous section, the condensation happens at
 the two commensurate momentum $ 0 $ and  $\mathbf{Q}=\mathbf{K}_2-\mathbf{K}_1=(\pi,0)$, so the order parameter  takes the form:
\begin{equation}
\langle \alpha_\mathbf{k}\rangle =\psi_1\delta_{\mathbf{k},0} +\psi_2\delta_{\mathbf{k},\mathbf{Q}}
\label{cantalpha}
\end{equation}
where $\psi_1, \psi_2$ are the two complex order parameters.

 One must use the Bogoliubov transformation Eq.\eqref{bogoliubovTrans} to establish the connection between the quantum spin
 and the two complex order parameters:
\begin{align}
	\langle S_i^+\rangle
	\propto
	u[\psi_1+(-1)^{i_x}\psi_2]
	+v[\psi_1^\ast-(-1)^{i_x}\psi_2^\ast]
	\propto
	(\psi_1+\psi_1^\ast)+(-1)^{i_x}(\psi_2-\psi_2^\ast)
	=\psi_R+(-1)^{i_x}i\psi_I
\label{compact}
\end{align}
where we have used the fact $u=u_{0}=u_{\mathbf{Q}}=\infty$
and $v=v_{0}=-v_{\mathbf{Q}}=\infty$, but their ratio $u/v=1$, so they can be factored out.
In fact, the Bogoliubov transformation matrix elements $ u, v $ are only finite at IC-momentum, but diverge at C-momentum.

Naively, similar to the last section, one may still need to use the two complex order parameters  $ \psi_1, \psi_2 $ to construct
the effective action. However,  Eq.\ref{compact} shows that the relevant order parameter maybe just ONE complex field as
$\psi=\psi_1+\psi_1^\ast+\psi_2-\psi_2^\ast$ whose real part $\psi_R=\Re \psi=\psi_1+\psi_1^\ast$ and
imaginary part $\psi_I=\Im \psi=-i(\psi_2-\psi_2^\ast)$ can be used to determine the quantum spin uniquely.
This observation is further substantiated by the crucial fact that under $ U(1)_{soc} $, $ \psi \rightarrow e^{ i \phi_0 } \psi $
as shown in the item 4 below. One may call this new phenomenon as order parameter reduction (OPR) from 2 to 1 which simplifies the following analysis considerably. Intuitively, one may also think $ \psi $ as a composite operator consisting of two components $ \psi_1, \psi_2 $,
one leads to its real part, the other leads to its imaginary part. The two components will emerge as two independent ones when getting into
a IC-phase. This fractionization  process indeed happens as shown in Sec V.

Because the Z-FM state breaks no symmetry of the Hamiltonian, so one can study how the single order parameter $\psi$
transform under symmetries of $\mathcal{H}$,
\begin{enumerate}
    \item Translation:
	    $\mathcal{T}_x: \psi(x,y)\to \psi^\ast(x,y)$
	    and $\mathcal{T}_y: \psi(x,y)\to \psi(x,y)$;
    \item Space reflection:
	    $\mathcal{I}_y: \psi(x,y)\to \psi(-x,y)$;
    \item Spin reflection:
	    $\mathcal{P}_z: \psi(x,y)\to -\psi(x,y)$;
    \item Spin-orbital reflection:
	    $\mathcal{T}\circ\mathcal{I}_x\circ\mathcal{P}_x:
	    \psi(x,y)\to -\psi^\ast(x,-y)$ and
	    $\mathcal{T}\circ\mathcal{I}_x\circ\mathcal{P}_y:
	    \psi(x,y)\to \psi^\ast(x,-y)$;
    \item Spin-orbital $ U(1)_{soc} $ rotation:
	    $\mathcal{R}: \psi(x,y)\to e^{i\phi_0}\psi(x,y)$;
    \item Enlarged mirror symmetry at $\beta=\pi/4$:
	    $\mathcal{T}\circ\mathcal{M}:
	    \psi(x,y)\to -\psi^\ast(x,y)$.
	    ( of course, $\beta=\pi/4$ is beyond this regime )
\end{enumerate}

The above symmetry analysis leads to
the following one complex component boosted effective action  with the dynamic exponent $z=1$ In the continuum limit:
\begin{align}
    \mathcal{S}
	=\int d\tau d^2r
	[(\partial_\tau\psi^*-ic\partial_y\psi^*)
	(\partial_\tau\psi-ic\partial_y\psi)
	+v_x^2|\partial_x\psi|^2
	+v_y^2|\partial_y\psi|^2
	-\mu|\psi|^2+U|\psi|^4]
\label{boostleft}
\end{align}
 where one need to realize
 $ (\partial_\tau\psi^*-ic\partial_y\psi^*)(\partial_\tau\psi-ic\partial_y\psi) \neq |(\partial_\tau\psi-ic\partial_y\psi)|^2 $.

Our microscopic calculation shows that $ \mu=h_{c2}-h, U > 0 $ and
$c\propto\sin2\beta$. Due to the magnon condensations at only the two C-momenta, the odd derivative in $\partial_y$ terms only appear
in the combination with $ \partial_{\tau} -ic \partial_y $ which specifies the kinetic term in Eq.\ref{boostleft}.

Note that  $ c=0 $ vanishes at the Abelian point $\beta=0$.
This is because that at the Abelian point $\beta=0$, in addition to the $ U(1)_{soc} $ symmetry, there is an enlarged space reflection $\mathcal{I}_x:\psi(x,y)\to \psi(x,-y)$.
See also appendix E.

\subsection{ The spin-orbital order of the ground state }

At the mean-field level, we can substitute $\psi\to\sqrt{\rho_0}e^{i\phi_0}$ into the effective action Eq.\ref{boostleft}
\begin{align}
    \mathcal{S}=-\mu\rho_0+U\rho_0^2
\end{align}
When $\mu= h_{c2}-h <0$, it is in the Z-FM state with $ \langle \psi \rangle =0$.
When $\mu>0$, it is in the canted phase with  $ \langle \psi \rangle=\sqrt{\rho_0}e^{i\phi_0}$
where $\rho_0=\mu/2U$ and $\phi_0$ is a arbitrary angle due to the U(1)$_\text{soc}$ symmetry.

Taking the real and imaginary part of  $ \langle \psi \rangle $, then combing Eq.\ref{compact} with  the constraint $|\mathbf{S}_i|^2=S^2$, one obtain
the spin-orbital order  of the canted phase as:
\begin{align}
    \langle S_i^+\rangle=\sqrt{\rho_0}[\cos\phi_0+(-1)^{i_x}i\sin\phi_0],
    \quad \langle S_i^z\rangle=\sqrt{S^2-\rho_0}
\label{cantorder}
\end{align}
where the sign of $\pm\sqrt{S^2-|S^+|^2}$ is
chosen such that $S_i^z$ reproduces the Z-FM order when $\rho_0\to0$.
It is obvious Eq.\eqref{cantorder} indeed matches the spin-orbital order of the canted phase
achieved by the microscopic SWE calculations in \cite{rhh}.
Remarkably, despite we only use one complex order parameter $ \psi $,
one can still use its real and imaginary part to stand for the transverse quantum spin with TWO
C- ordering wavevectors $ (0,0) $ and  $\mathbf{Q}=\mathbf{K}_2-\mathbf{K}_1=(\pi,0)$.

\subsection{ Excitation spectrum: Exotic Goldstone mode and Higgs mode }

   In the Z-FM phase, $\mu < 0$, one can write $\psi=\psi_R+i \psi_I$ as its real part and imaginary part
and expand the action upto second order
\begin{align}
    \mathcal{S}
	=\int d\tau d^2r
	\sum_{\alpha=R,I}
	[(\partial_\tau \psi_\alpha
	-ic\partial_y \psi_\alpha)^2
	+v_x^2(\partial_x  \psi_\alpha)^2
	+v_y^2(\partial_y   \psi_\alpha)^2
	-\mu( \psi_\alpha)^2]
\end{align}
which lead to 2 degenerate gapped modes
\begin{align}
	\omega_{R,I}=\sqrt{-\mu+v_x^2k_x^2+v_y^2k_y^2}-ck_y
\label{gapleft}
\end{align}
which match the results achieved by SWE in \cite{rhh}.
Eq.\ref{gapleft} can be contrasted to Eq.\ref{gapmiddle}, both are gapped modes in the Z-x phase.
The difference is that the latter is expanded around the two true in-commensurate minima $\mathbf{K}_1=(0,k_0), \mathbf{K}_2=(\pi,k_0)$
whose constant contour is shown in Fig.1b and indicates the dynamic exponent $ z=2 $
while the former is expanded around the two commensurate momentum $ (0,0) $ and $ (\pi,0) $ which are not the true minima
until hitting the left segment of $ h_{c2} $
as shown in Fig.1b, it indicates the dynamic exponent $ z=1 $.

In the canted phase, $\mu>0$, we can write the fluctuations in the polar coordinates
$\psi=\sqrt{\rho_0+\delta\rho}e^{i(\phi_0+\delta\phi)}$
and expand the action up to the second order in the fluctuations:
\begin{eqnarray}
    \mathcal{S}
	& = & \frac{1}{2\rho_0}\int d\tau d^2r\Big(
	[(\partial_\tau-ic\partial_y)\delta\rho]^2
	+[v_x^2(\partial_x\delta\rho_+)+v_y^2(\partial_y\delta\rho)]
	+4\rho_0U(\delta\rho)^2     \\  \nonumber
	& +  &\rho_0^2[(\partial_\tau+ic\partial_y)\delta\phi]^2
	+\rho_0^2[v_x^2(\partial_x\delta\phi)^2
		 +v_y^2(\partial_y\delta\phi)^2]
	\Big)
\label{GoldHiggsact}
\end{eqnarray}
 which due to $ z=1 $, leads to one gapless Goldstone mode and one gapped Higgs mode \cite{cavity}
\begin{eqnarray}
    \omega_{\text{H}}
	& = & \sqrt{4\rho_0U+v_x^2k_x^2+v_y^2k_y^2}-c k_y    \\  \nonumber
    \omega_{\text{G}}
	& = & \sqrt{v_x^2k_x^2+v_y^2k_y^2}-c k_y
\label{GoldHiggs}
\end{eqnarray}
where the Goldstone mode reproduce the superfluid mode
and the Higgs mode reproduce the "roton" mode achieved by SWE in \cite{rhh}.

Note that it is the $ z=1 $ which ensures the separation of the real part from the imaginary part when $ \mu <0 $ in the Z-x phase in Eq.\ref{gapleft} and the separation of the Higgs mode from the Goldstone mode when $ \mu > 0 $ in the canted phase in  Eq.\ref{GoldHiggs}.
Intuitively, one can say the two degenerate gapped modes in Eq.\ref{gapleft} turn into the Goldstone mode and the Higgs mode in Eq.\ref{GoldHiggs}
through the QPT from the Z-x phase to the canted phase at $ h_{c2} $.

\subsection{The QCP: a boosted SF-Mott transition }

If putting $ c=0 $ in the effective action Eq.\ref{boostleft}, it is nothing but a 3D XY universality class which respects the
Lorentz invariance. If $ c > 0 $, it can be transformed back into a 3D XY universality class in a boosted frame along $ y- $ axis by
performing a Galileo transformation $ y^{\prime}= y- ct, t^{\prime}=t $.
In the imaginary time $ \tau= it $, it implies $ \partial^{\prime}_y \to \partial_y,
\partial^{\prime}_\tau \to \partial_\tau -i c \partial_y $. So the effective action becomes the same as
the 2d SF-Mott transition with $ z=1 $ in a boosted frame.
However, the action at $ c=0 $ is Lorentz invariant instead of  Galileo invariant, so the Galileo  boost must lead to some dramatic effects.
Indeed, as to be discussed in Sec.V,  it is the boost which drives the quantum Lifshitz transition from
the canted phase to the IC-SkX phase at $ \beta=\beta_L $.

The $ z=1 $ is protected by the Lorentz invariance at $ c=0 $.
Any $ c > 0 $ breaks Lorentz invariance. So the action is neither Lorentz invariant nor Galileo invariant.
The mechanism for how a SOC generates such a boost is not known and need to be investigated further.
The $ c $ term is marginal at $ h_{c2} $ suggesting a line of fixed points.
The interaction $ U $ term is marginally irrelevant at $ c=0 $.
How does the fact change along the fixed line need to be determined by RG calculations.
If the dynamic exponent $ z=1 $ receives anomalous dimension need to be examined also \cite{S:un}.

At the Abelian point $ \beta=0 $, $ c=0 $, the boost disappears, so the transition at $ h_{c2} $ is nothing but
the 3d $XY$ universality class. In fact, as shown in \cite{rh,rhh,rhtran}, the Hamiltonian at this Abelian point
can be mapped to a FM Heisenberg model in a staggered Zeeman field along $ x-$ direction. As shown in the appendix F,
it is dramatically different than the AFM in a uniform field which has the dynamic exponent $ z=2 $.

\subsection{ Contrast to a putative supersolid }

 In the previous works on putative supersolids in a continuum system driven by the roton collapse \cite{SS1,SS2,SS3,SSrev},
 There is a  crucial coupling term which couples the lattice phonon modes to the SF mode.
 $ i a_{\alpha \beta} u_{\alpha \beta}\partial_{\tau} \theta $  where
 $ u_{\alpha \beta}= \frac{1}{2} ( \partial_{\alpha} u_{\beta} + \partial_{\beta} u_{\alpha} ) $  is the linearized strain tensor.
 The factor of $ i $ is important in this coupling. By integration by parts, this term can also be
 written as $ a_{\alpha \beta} (\partial_{\tau} u_{\beta} \partial_{\alpha} \theta
 + \partial_{\tau} u_{\alpha} \partial_{\beta} \theta   )  $   which has the clear
 physical meaning of the coupling between the SF velocity
 $ \partial_{\alpha} \theta $ and the velocity of the lattice vibration $ \partial_{\tau} u_{\beta} $. It is
 this coupling between the phonon mode  and the superfluid mode which leads to
 the two gapless low energy modes inside the SS. They have
 their own characteristics which could be detected by experiments. The two gapless modes result from
 $ U(1)_c \times U(1)_l  \to 1 $ symmetry breaking, the first is the phase, the second the lattice translational symmetry breaking.
 In a contrast, the coset Eq.\ref{coset} only leads to one gapless mode and one roton mode.
 So the second term $ -i2c \partial_{\tau} \psi^{*} \partial_y \psi $ in Eq.\ref{boostleftM} is very similar to
 such a coupling in the putative supersolid.

\section{Quantum Lifshitz transition at the left critical field $\beta_{L}$: Order parameter fractionization}

 Inside the canted phase at a fixed $ h $, as the SOC parameter increases,
 there is a quantum Lifshitz  transition from the canted phase to the IC-SkX driven by the instability
 of the Goldstone mode in Eq.\ref{GoldHiggsact} ( Fig.1 ). Because the gapped Higgs mode remains un-critical across
 the transition, one can simply drop it. Although the Goldstone mode to the quadratic order in Eq.\ref{GoldHiggsact} is enough
 inside the canted phase. When studying  the transition to the IC-SkX, one must incorporate higher derivative terms and
 also higher order terms to the Goldstone mode in Eq.\ref{GoldHiggsact}. A simple symmetry analysis leads to the following   bosonic quantum Lifshitz transition at the left critical
 SOC parameter $\beta_L$ (Fig.1) which extends Eq.\ref{GoldHiggsact} to include higher derivative terms and
 also higher order terms:
\begin{equation}
    \mathcal{S}_L
	=\int d\tau d^2r
	[(\partial_\tau\phi-ic\partial_y\phi)^2
	+v_x^2(\partial_x\phi)^2
	+v_y^2(\partial_y\phi)^2
	+a(\partial_y^2\phi)^2
	+b(\partial_y\phi)^4]
\label{boosttoright}
\end{equation}
where $a,b>0$ and $c\propto\sin(2\beta)$, especially $ v^2_y- c^2 =\beta-\beta_L $ is the tuning parameter.
At a fixed $ h $, as $ \beta $ increases, the boost $ c $ also increases. When $ c $ reaches the value of $ v_y $,
it signifies an instability of the Goldstone mode which drives the quantum Lifshitz transition from the canted phase to the IC-SkX phase. \cite{morederivative}. A simple scaling shows that when $ z=1 $ inside the canted phase $ [a]=-2, [b]=-3 $, so they are irrelevant
inside the canted phase, but become important near the transition as to be shown in the following.

\subsection{ Obtain the spin-orbital order of the IC-SkX from the canted phase: Order parameter fractionization }

The mean-field state can be written as $\phi=\phi_0+k_0 y$.
Substituting it to the effective action Eq.\ref{boosttoright}, we obtain
\begin{align}
    \mathcal{S}_{0}
	\propto(v_y^2-c^2)k_0^2+bk_0^4
\end{align}

At a lower boost $c^2 < v_y^2 $, $k_0=0$ is in the C- Canted phase.

At a high boost  $ c^2 > v_y^2 $, $k_0^2=(c^2-v_y^2)/2b$ is inside the IC-SkX phase with the modulation $ k_0 $ along the $ y-$ axis.
The sign of $k_0$ is determined by the sign of $c$, i.e. $k_0=\mathrm{sgn}(c)\sqrt{k_0^2}$.
Substituting $\phi=\phi_0+k_0y$ back to the phase of the complex order parameter leads to
$ \psi=\psi_1+\psi_1^\ast+\psi_2-\psi_2^\ast = \tilde{\psi} e^{i(\phi_0+k_0y)}$,
which admits a physical solution \cite{admit} with
$\psi_\alpha=\tilde{\psi}_\alpha e^{ik_0y}, \alpha=1,2 $.
Thus Eq.\ref{compact} in the canted phase turns into:
\begin{align}
    \langle S_i^+\rangle
	=u[\tilde{\psi}_1+(-1)^{i_x}\tilde{\psi}_2]e^{ik_0i_y}
	+v[\tilde{\psi}_1^\ast-(-1)^{i_x}\tilde{\psi}_2^\ast]e^{-ik_0i_y}
\end{align}
 where we put back the two phenomenological parameters $ u $ and $ v $.
 This is because  $u/v\neq 1$ any more due to a nonzero $k_0$. Thus it reproduces the IC-SkX phase in Eq.\ref{pmk0y} when $c^2>v_y^2$.
This is equivalent to shift the two condensation wave-vectors in Eq.\ref{cantalpha} to
$\langle\alpha_\mathbf{k}\rangle
=\tilde{\psi}_1\delta_{\mathbf{k},0+(0,k_0)}
+\tilde{\psi}_2\delta_{\mathbf{k},\mathbf{Q}+(0,k_0)}$ at the very beginning.


 So in the C-IC quantum Lifshitz transition from the canted phase to the IC-SkX phase,
 the order parameter fractionize from One complex order parameter $ \psi =\psi_1+\psi_1^\ast+\psi_2-\psi_2^\ast $  into
 TWO independent ones $ \psi_1, \psi_2 $. This fractionization \cite{admit} is caused by the appearance of the IC- \cite{notOPF}.
 One may also look at the quantum Lifshitz transition from the dual point of view:
 there is a IC-C transition from the IC-SkX phase to the canted phase,
 the TWO complex order parameters $ \psi_1, \psi_2 $ confine into just One complex order parameter
 $ \psi= \psi_1+\psi_1^\ast+\psi_2-\psi_2^\ast $.
 The dynamic exponent changes from $ z=1 $ to $ z=2 $, the Higgs mode in Eq.\ref{GoldHiggs} in the canted phase
 automatically changes to the roton mode in Eq.\ref{GoldRoton} inside the IC-SKX phase.
 In fact, the order parameter fractionization (OPF) already shows its sign even above the $ h_{c2} $
 inside the Z-FM phase: Eq.\ref{gapleft} containing 2 degenerate gapped modes with real and imaginary part
 above canted phase evolve into  Eq.\ref{gapmiddle} containing the 2 degenerate gapped modes with two complex order parameters
 above the IC-SkX phase.

\subsection{ The excitation spectrum in the canted phase and IC-SkX phase  }

At a low boost $c^2<v_y^2$, the quantum phase fluctuation can be written as $\phi=\phi_0+\delta\phi$.
Expanding the action upto second order  leads to:
\begin{align}
    \mathcal{S}_{2c}
	=\int d\tau d^2r
	[(\partial_\tau\phi-ic\partial_y\phi)^2
	+v_x^2(\partial_x\phi)^2
	+v_y^2(\partial_y\phi)^2]
\label{s2cant}
\end{align}
   which reproduces the gapless Goldstone mode in Eq.\ref{GoldHiggs} inside the canted phase:
\begin{align}
    \omega_\mathbf{k}
	=\sqrt{v_x^2 k_x^2+v_y^2k_y^2}-ck_y
\end{align}

 At a high boost $ c^2>v_y^2 $, the quantum phase fluctuations can be written as $ \phi=\phi_0+k_0y+\delta\phi $.
 Expanding the action upto the second order in the phase fluctuations leads to
\begin{align}
    \mathcal{S}_{2ic}
	=\int d\tau d^2r
	[(\partial_\tau\phi-ic\partial_y\phi)^2
	+v_x^2(\partial_x\phi)^2
	+(v_y^2+6bk_0^2)(\partial_y\phi)^2]
\label{s2ic}
\end{align}
   which reproduces the gapless Goldstone mode in Eq.\ref{GoldRoton} or Eq.\ref{Goldhc1} inside the IC-SkX phase:
\begin{align}
    \omega_\mathbf{k}
	&=\sqrt{v_x^2 k_x^2+(v_y^2+6bk_0^2)k_y^2+ak_y^4}-ck_y   \\  \nonumber
	&=\sqrt{v_x^2 k_x^2+(3c^2-2v_y^2)k_y^2+ak_y^4}-ck_y
\end{align}
where one can see  $3c^2-2v_y^2>= 2(c^2-v_y^2)+ c^2 > c^2 $ when $c^2>v_y^2$,
thus the $\omega_\mathbf{k}$ is stable in IC-SkX phase.

\subsection{The exotic QCP scaling with the dynamic exponents $ (z_x=3/2, z_y=3 )  $ }

    It is instructive to expand the first kinetic term in Eq.\ref{boosttoright}  as:
\begin{align}
    \mathcal{S}
	=\int d\tau d^2r
	[Z(\partial_\tau\phi)^2
	-2iv_y\partial_\tau\phi\partial_y\phi
	+v_x^2(\partial_x\phi)^2 + \gamma (\partial_y \phi)^2
	+a(\partial_y^2\phi)^2
	+b(\partial_y\phi)^4]
\label{ab}
\end{align}
  where $ Z $ is introduced to keep track of the renormalization of  $ (\partial_\tau\phi)^2 $ and
  $ \gamma= v^2_y- c^2 =\beta- \beta_L $ is the tuning parameter.

  The scaling $ \omega \sim k^3_y, k_x \sim k^2_y $ leads to the exotic the dynamic exponents $ (z_x=3/2, z_y=3 )  $.
  Then one can get the scaling dimension of $ [\gamma]=2 $ which is relevant, as expected, to tune the transition.
  One can also find that $ [Z]=[b]=-2 < 0 $, so both are leading irrelevant operators\cite{morederivative} which
  determine the finite $ T $ bahaviours ( see Sec.VI-3 ）.
  Setting $ Z=b=0 $ in Eq.\ref{ab} leads to the fixed action
  at the QCP where $ \gamma=0 $. It is instructive to compare Eq.\ref{ab} with Rokhsar-Kivelson's Quantum Dimer (QD) model in a square lattice
  in its height representation \cite{dimer,dimer1,dimer2}
\begin{eqnarray}
{\cal L}_{QD} & = & \kappa ( \partial_\tau \chi )^2 + \rho_s ( \nabla \chi)^2 + K ( \nabla^2 \chi)^2
       +  u ( \nabla \chi)^4     \nonumber  \\
       & + & \lambda \cos 2 \pi\chi   + \cdots
\label{qd}
\end{eqnarray}
   At the QCP $  \rho_s=0 $, there is a line of fixed point controlled by the parameter $ K $ with the dynamic exponent $ z=2 $
   describing the transitions between various VBS.

  The main differences are (1) The monopole term $ \lambda \cos 2 \pi \phi $ is absent in Eq.\ref{ab}.
  While the boost term $ -2iv_y\partial_\tau\phi\partial_y\phi $ is absent in the QD model Eq.\ref{qd}.
  (2) Here, the dynamic exponent is anisotropic with $ (z_x=3/2, z_y=3 )  $ due to the boost term, while that $ z=2 $ is
  isotropic in the QD model. (3) Of course, our system is a quantum spin one.
  (3) Because $ U(1)_{soc} $ is broken in both the canted phase and IC-SkX phase, the phase windings or vortex excitations in $ \phi $ may not be important in Eq.\ref{ab}. But it is important in Eq.\ref{qd} encoded in the monopole term.
  This monopole term is Type-I dangerously irrelevant near the RK point, it sets up the periodicity of $ \chi \rightarrow \chi + 1 $,
  so it is responsible for various VBS
  and also possible in-complete devil staircases of all the in-commensurate VBS phases in the tilted side $ \rho_s < 0 $.
  For both complete and in-complete devil staircases at a generic $ (\alpha, \beta ) $ and also more contrasts with the QD model from
  different perspectives, but no Zeeman field, see  \cite{devil}.

\subsection{ Multi-critical point M }

 Expanding the kinetic term in Eq.\ref{boostleft} leads to
\begin{equation}
    \mathcal{S}
	=\int d\tau d^2r
	[ Z | \partial_\tau \psi |^2 -i2c \partial_{\tau} \psi^{*} \partial_y \psi
	+v_x^2|\partial_x\psi|^2+ a |\partial^2_y\psi|^2	+\gamma |\partial_y\psi|^2	-\mu|\psi|^2+U|\psi|^4]
\label{boostleftM}
\end{equation}
 where $ \mu= h_{c2}-h $ and $ \gamma= v^2_y-c^2 $.
 Moving along $ h_{c2} $ in the left range $ 0 < \beta < \beta_1 $, $ \gamma $ decreases until reaching the Multi-critical (M) point $ \gamma =0 $. So there are two relevant operators $ \mu $ and
 $ \gamma $, with the scaling dimensions $ [\gamma]=2, [\mu]=4 $ respectively.
 Then there is a order parameter fractionization (OPF) at the $ M $ point:
 one complex order parameter $ \psi $ splits into $ \psi_1, \psi_2 $, C to IC transition,
 dynamic exponent changes from $ z=1 $ to $ z=2 $ through the M point with $ (z_x=3/2, z_y=3 )
 $.

\section{ Finite Temperature phase transitions and quantum critical regimes }

 Any experiments are performed at finite temperatures which are controlled by the quantum phases and phase transitions at $ T=0 $ in Fig.1 and Fig.2. The experiments in \cite{halfinteger,unquantized} examined carefully the interplay of the temperature against the Zeeman field.
 Here, we discuss the effects of finite temperatures.
 The thermodynamic quantities at a small finite $ T $ was discussed in \cite{rhh}. Here,
 we focus on the spin-spin correlation functions at a finite $ T $.


\begin{figure}[!htb]
\includegraphics[width=10cm]{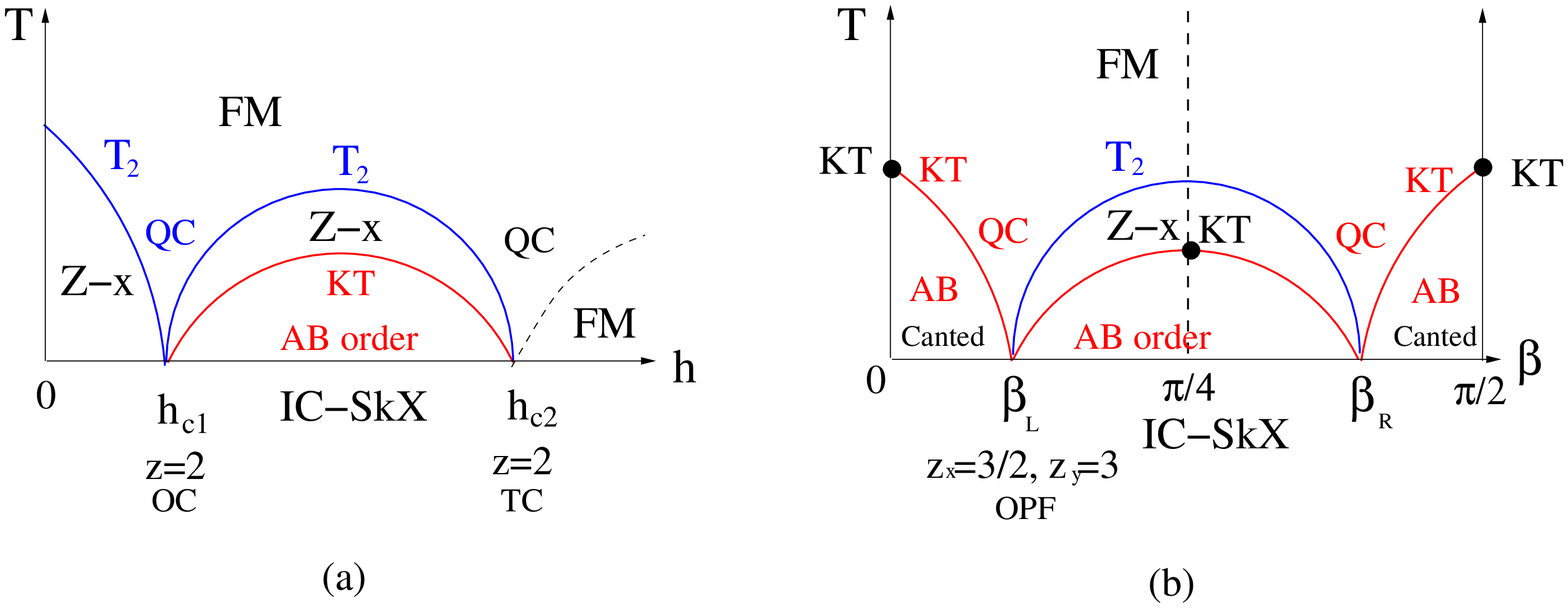}
\hspace{0.5cm}
\includegraphics[width=4.5cm]{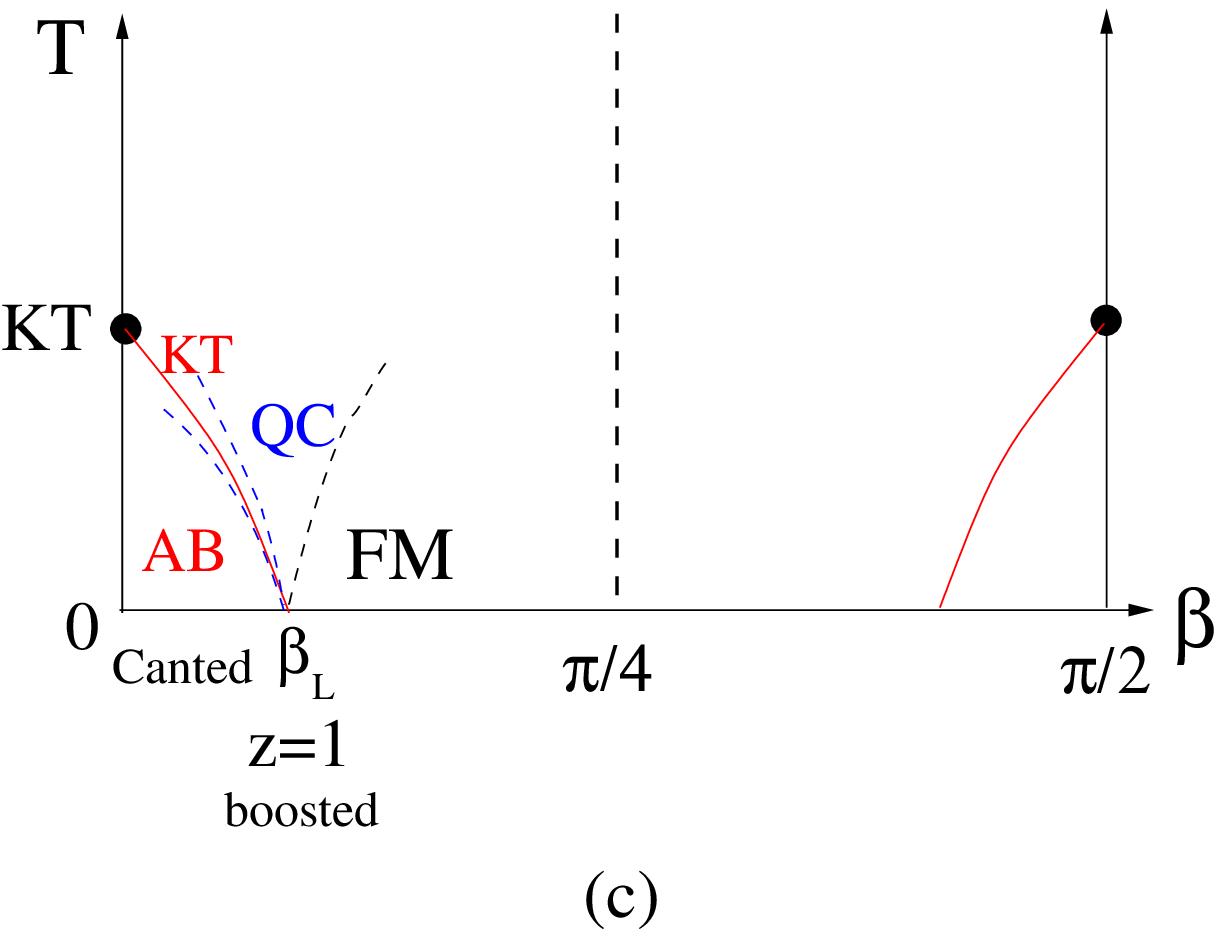}
\caption{  (Color online) Finite temperature phase transitions above the three quantum C-IC transition at $ T=0 $.
The zero temperature QPT with various dynamic exponents and associated QC regimes are also indicated.
OC means one component $ z=2 $ with one type-II dangerously operators.
TC means Two component $ z=2 $ with TWO type-II dangerously operators. OPF means the order parameter fractionization.
(a) At a fixed $ \beta $. At $ T=0 $, there is a quantum C-IC transition  from the Z-x to the IC-SkX at $ h=h_{c1} $ and
from the IC-SkX to the FM at $ h=h_{c2} $ shown in Fig.1. There is a finite temperature Ising transition $ T_2 $ above the Z-x state.
The IC-SkX has only an algebraic  ( denoted as AB in the figure ) order in the transverse spin components before
getting to the $ Z-x $ state at $ T=T_{KT} $, then melt into the FM state at $ T_2 $.
As shown in the text, the transition at $ T_{KT} $ is the same universality class as the Koterlitz-Thouless (KT) transition,
even away from the mirror symmetric point $ \beta=\pi/4 $ where $ T_{KT} $ reaches the maximum value.
as shown as the black dot in (b).
(b) At a fixed $ h $.  At $ T=0 $, there is a quantum C-IC transition  from the canted phase to the IC-SkX at $ \beta_L $ and
from the IC-SkX to the mirror reflected canted phase at $ \beta_R=\pi/2-\beta_L $ shown in Fig.1.
There is a finite temperature KT transition above the canted state, even away from the two Abelian points $ \beta=0, \pi/2 $
where $ T_{KT} $ reaches the maximum. There is a mirror symmetry about $ \beta=\pi/4 $ where the IC-SkX reduces to the $ 2 \times 4 $ SkX and the $ T_{KT} $ reaches the maximum.
Replacing the IC-SkX in (b) by the FM leads to (c) where there is a C-C transition  from the canted to the FM state at $ T=0 $ in Fig.1.
As argued in \cite{rhh}, all the critical temperatures $ T_c \sim  \Delta \sim 2S J  = NJ \sim  N \times 0.2 nK  $ where the $ N $ is the number of atoms per site, so all the critical temperatures can be easily increased above the experimentally reachable temperatures simply by increasing the number of spinor atoms on every lattice site. Fig.2a can be contrasted to the experimental temperature versus Zeeman field phase diagrams in  \cite{halfinteger,unquantized}. }
\label{htzx}
\end{figure}

   As argued in \cite{rh}, there is only one finite temperature phase transition in the Ising universality class \cite{rhtran} above the $ Z-x $  phase. The FM state breaks no symmetries of the Hamiltonian, so no transitions above it. So we only need to discuss
   the finite temperature transitions above the canted phase and IC-SkX state as shown in Fig.\ref{htzx}a.

At a finite temperature, setting the quantum fluctuations ( the $ \partial_{\tau} $ term ) vanishing, in Eq.\ref{s2cant} or Eq.\ref{s2ic},
 then both equations reduce to
 \begin{align}
    \mathcal{S}_{KT}
	=\int d^2r
	[ v_x^2(\partial_x\phi)^2 + \gamma (\partial_y\phi)^2]
\label{KT}
\end{align}
 where $ \gamma= v^2_y- c^2 $ inside the canted phase and  $ \gamma=2(c^2-v_y^2) $ inside the IC-SkX phase.
 It indicates the finite temperature phase transition is still in Kosterlize-Thouless (KT) universality class, despite the exotic form of the
 spectrum of the Goldstone mode.


{\sl 1. The canted phases: }
   In the canted phase, from \ref{s2cant}, one can see that at any $ T>0 $, the Goldstone mode fluctuations
   Eq.\ref{GoldHiggs} lead to $ \langle S^{+} \rangle=0 $ in Eq.\ref{cantorder},
   so the transverse spin correlation functions display algebraic orders at the two ordering wavevectors $ \vec{Q}_1=(0,0) $ and $ \vec{Q}_2=(\pi,0) $.
   So there is only one finite temperature phase transition $ T_{KT} $ driven by the topological defects
   in the phase $ \phi $ in Eq.\ref{KT} above the canted phase to destroy the algebraic order ( Fig.2b,c ).

   The transverse Bragg spectroscopy in the canted phase at $ T=0 $ will display sharp peaks at  $ \vec{Q}_1=(0,0) $ and $ \vec{Q}_2=(\pi,0) $.
   However at $ 0 < T < T_{KT} $, the transverse peaks at  $ \vec{Q}_1 $ and $ \vec{Q}_2 $ will be  replaced by some power law singularities \cite{socsdw}.  At $  T > T_{KT} $, the power law singularities disappear.

{\sl 2. The IC-SkX phase: }

   In the IC-SkX phase, from \ref{s2cant}, one can see that at any $ T>0 $, the Goldstone mode fluctuations Eqn.\ref{Goldhc1}
   ( or Eqn.\ref{GoldRoton}  also lead to $ \langle S^{+} \rangle=0 $ in Eq.\ref{orderuv} ( or Eq.\ref{ordercs} ),
   so the transverse spin correlation functions also display algebraic orders at the four
   in-commensurate ordering wavevectors  $ (0, \pm k^{0}_y ) $ and $ (\pi, \pm k^{0}_y ) $.
   So there are two finite temperature phase transitions above the IC-SkX state:
   one transition $ T_{KT} $ in the transverse spin sector to destroy the algebraic order, then
   another Ising  $ Z_2 $  transition in the longitudinal spin sector $ T_2 $ to destroy the $ A $ and $ B $ sublattice $ Z_2 $
   symmetry breaking as shown in Fig.1a.
   We also expect $ T_{KT} < T_{2} $. Of course, at all the quantum phase transition boundaries in Fig.1, $ T_{KT}=T_{2}=0 $.



   The elastic longitudinal Bragg spectroscopy in the IC-SkX at $ T=0 $ will display a sharp peak at  $ (\pi,0) $,
   while the transverse Bragg spectroscopy will display sharp peaks at  the four
   in-commensurate ordering wavevectors  $ (0, \pm k^{0}_y ) $ and $ (\pi, \pm k^{0}_y ) $.
   However at $ 0 < T < T_{KT} $, the transverse peaks at  $ (0, \pm k^{0}_y ) $ and $ (\pi, \pm k^{0}_y ) $ will be
   replaced by  some power law singularities \cite{socsdw},
   the longitudinal peak remains sharp. At $ T_{KT} < T < T_{2} $, the power law singularities disappear, but the longitudinal peak remains sharp.
   When $ T > T_2 $, the longitudinal peak disappears.

  Following the procedures \cite{scaling,tqpt}, one can also derive the scaling functions of spin-spin correlation functions
  at finite temperatures across the three C-IC quantum transitions in Fig.\ref{htzx}a,b and also the C-C transition from the canted phase to the FM at the left  or right segment of $ h_{c2} $ in Fig.2c.

{\sl 3. The quantum critical regimes }

 The QC scaling at $ h_{c1} $ with $ z=2 $ in Fig.2a was derived in \cite{z2}.
 The one type-II dangerously irrelevant operator $ V $ will not affect
 the leading order scalings. Unfortunately, the universality class at $ T=0, h= h_{c2} $  with $ z=2 $  in Fig.2a is still un-known, so the QC scaling remains to be determined. The QC scaling at $ \beta_L $  with $ z=(3/2,3) $
 in Fig.2b can be derived by using the two leading irrelevant operators $ Z $ and $ b $
 in the effective action Eq.\ref{ab}.  Unfortunately, the universality class at $ T=0, \beta=\beta_L $  in Fig.2c is still un-known, so the QC scaling remains to be determined.

\section{ Implications to materials with strong SOC in a Zeeman field }

  Although the RFHM was derived as the strong coupling model of interacting spinor boson Hubbard model at integer fillings in the presence of SOC,
  we may just treat it as an effective lattice quantum spin model which incorporate competitions among
  Heisenberg term, Kitaev term and Dzyaloshinskii-Moriya ( DM ) term. As shown in \cite{rh}, when expanding the two $R$ matrices in Eqn.\ref{rhgeneral},  one can see that it leads to
a Heisenberg + Kitaev  ( strictly speaking, the quantum compass model in a square lattice ) + Dzyaloshinskii-Moriya (DM) interaction
\begin{equation}
 H_{s}=-J [\sum_{\langle i j \rangle  } J^a_{H} \vec{S}_{i} \cdot \vec{S}_{j}
	+\sum_{\langle i j \rangle a } J^{a}_{K} S^{a}_{i} S^{a}_{j}
	+\sum_{\langle i j \rangle a } J^{a}_{D} \hat{a} \cdot \vec{S}_{i} \times \vec{S}_{j} ]
\end{equation}
where $ \hat{a}= \hat{x}, \hat{y} $, $ J^{x}_H=\cos 2 \alpha,  J^{y}_H=\cos 2 \beta $;
$ J^{x}_K= 2 \sin^2 \alpha,  J^{y}_K= 2 \sin^2 \beta $ and $ J^{x}_D=\sin 2 \alpha,  J^{y}_D=\sin 2 \beta $.

Along the whole solvable line  $ (\alpha=\pi/2, \beta ) $, we can write:
\begin{equation}
 J^{x}_H=-1,  J^{y}_H=\cos 2 \beta;~~~~~
  J^{x}_K= 2,  J^{y}_K= 2 \sin^2 \beta;~~~~~
  J^{x}_D=0,  J^{y}_D=\sin 2 \beta
\end{equation}
  It is easy to see $  J^{y}_H > 0 $ when $ \beta < \pi/4 $, $  J^{y}_H < 0 $ when $ \beta > \pi/4 $ and vanishes at $ \beta=\pi/4 $.
  When $ \beta > \pi/4 $, the FM Kitaev term dominates, plus a AFM Heisenberg term in both bonds, plus a DM term in XZ plane
  $  J \sin 2 \beta ( S_{ix}S_{jz}- S_{iz}S_{jx}) $.
  So the RFHM  could be an alternative to the minimal $ (J,K,I) $ model used in \cite{kitaevlattice}
  or to the minimal $ (J,K, \Gamma ) $ model used in \cite{kim} to fit the experimental data phenomenologically.
  One common thing among all the three models is that it is dominated by FM Kitaev term, plus a small AFM Heisenberg term.
  The difference comes from the third term which, in our model is the crucial DM term.
  The Zeeman field adds a new dimension to these competitions which lead to the IC-SkX state in the center regime in Fig.1 and Fig.2.
  So RFHM + H can be used to not only to describe cold atom systems as described in details in \cite{rhh},
  but also the universal features of some strongly correlated materials which host some of these interactions.

The IC-SkX phase in Fig.\ref{globalphase} can be realized in some materials with a strong Dzyaloshinskii-Moriya (DM) interaction.
Indeed, a 2D skyrmion lattice has been observed between $ h_{c1}=50$ mT  and $ h_{c2}=70$ mT
in some chiral magnets \cite{sky4} MnSi or a thin film of Fe$_{0.5}$Co$_{0.5}$Si \cite{sky4}.
The effective actions Eq.\ref{z2single} and \ref{pmbasis} or \ref{pmbasisM} may be used to describe
the transitions near $ h_{c1} $ and $ h_{c2} $.

Recently, there are flurries of theoretical and experimental researches  to investigate the response of so called
Kitaev materials to a Zeeman field. For example, in the 4d Kitaev material $ \alpha-Ru Cl_3 $, the ground state was shown
experimentally  to have a Zig-Zag order.
In the application of a parallel magnetic field to the Zig-Zag magnetization \cite{angleH} at temperature as low to 2 $ K $,
the system stays in the Zig-Zag order upto a lower critical field $ \mu_0 h_{c1} \sim 7 $ T,
becomes fully polarized above a upper critical field $ \mu_0 h_{c2} \sim 9 $ T. Most interestingly,
in the intermediate field range  $  h_{c1} < \mu_0 H^{*}_{l} < h_{c2} $, there is a possible field-induced quantum spin liquid (QSL) ground state
displaying half-integer quantized thermal Hall conductivity plateau \cite{halfinteger} similar to those discovered in Fractional quantum Hall systems near $ \nu=5/2 $ \cite{NonFQHE}.
It hints a topologically protected chiral  Majorana fermion edge mode.
This edge mode is a direct consequence of the bulk Ising non-Abelian anyons in the Kitaev honeycomb lattice model subject to
a small Zeeman field along $ [111] $ direction .
The thermal Hall conductivity measurements in \cite{unquantized} between the ordering temperature of the Zig-Zag phase at $ T_N\sim 7 $ K and
the characteristic temperature of the Kitaev interaction $ J_K/k_B \sim 80 $ K also shows signatures compatible  with the itinerant Majorana fermions.
This exciting, although still controversial discovery inspires further experimental and  theoretical investigations.
For example, by performing un-controlled parton construction mean field theory, the authors in \cite{spinon} suggested that when the Kitaev model subjects to a Zeeman field along $ [111] $ direction, there could be a intermediate gapless  $ U(1) $ QSL phase at
an $ h_{c1} < h < h_{c2} $ with spinon Fermi surface
which shows un-quantized thermal Hall conductivity.
They also argued that the topological transitions at $ h_{c1} $ and $ h_{c2} $
are similar to the transition from a weak BCS pairing $ p_x+i p_y $ superconductor to a metal, then to a
band insulator respectively. They also alerted to the readers that the gauge field fluctuations may be ignored in the
gapped non-Abelian phase, but may be important in the  gapless  $ U(1) $ QSL phase, but very difficult to handle in a controlled way.
Unfortunately, despite many appealing theoretical proposals summarized in \cite{spinon},  there is not a consistent and coherent theoretical framework
which puts the Zig-Zag phase, the bulk gapped Kitaev non-abelian spin liquid phase with the half-integer quantized thermal Hall conductivity and
the gapless putative spinon Fermi surface with an un-quantized thermal Hall conductivity in
the same temperature versus the parallel Zeeman field phase diagram.
Obviously, despite the Zig-Zag phase is a magnetic ordered  ( therefore boring ) phase, it is
the parent state, takes a large portion of the phase diagram, can not be ignored in giving a
consistent description of experimental data in $ \alpha-Ru Cl_3 $.

Here, instead of directly working on Kitaev honeycomb lattice model,
we take an alternative approach to study the interplay of the Zeeman field and SOC in a different strongly correlated quantum spin model called Rotated Heisenberg model along its solvable line with the $ U(1)_{soc} $ symmetry \cite{rh,rhh,rhtran}.
The global phase diagram Fig.1 achieved by both controlled microscopic SWE and symmetry based phenomenological effective actions
is on a square lattice, so not directly relevant to the current experiments \cite{halfinteger} yet.
However, it does gives some physics universal to the competitions among
AFM Heisenberg interaction, FM Kitaev interaction,  DM  term and the Zeeman term.
For example, it indicates the interplay does lead to a highly non-trivial
intermediate phase sandwiched between the magnetic ordered phase below the low critical field $ h < h_{c1} $
and the fully polarized FM phase above the upper critical field $ h > h_{c2} $. Here, the intermediate phase
is the canted phase at a small SOC and the IC-SkX phase  at a large SOC.
The Z-x phase in a square lattice at a low field maybe used to mimic the Zig-Zag phase at a low field in a honeycomb lattice.
Of course, the Z-FM phase at a high field always exists in any case.
The IC-SkX phase could be easily melt into a QSL under some further quantum fluctuations. An extra SOC parameter in a honeycomb lattice
may provide such quantum fluctuations.
A future study on a honeycomb lattice with either spinor bosons or fermions \cite{rafhm} could be directly relevant.
If the IC-SkX indeed melts into a QSL, then the transitions at $ h_{c1} $ and $ h_{c2} $ will become
two Topological transitions driven by the condensation of spinons or $ Z_2 $ flux or some fermions instead of some order parameter condensations.

{\sl 1. The thermal Hall conductivities in all the phases in Fig.1 and Fig.2 }

Here, the thermal carriers are bosons instead of fermions in \cite{spinon}.
As argued in \cite{thermalhigh}, one needs to break both Time reversal and the particle-hole (PH) symmetry to get a non-vanishing thermal Hall
conductivities.
We expect that the extra term of the gapless Goldstone mode Eq.\ref{GoldHiggs} in the canted phase and Eq.\ref{GoldRoton} ( or Eq.\ref{Goldhc1} ) inside the canted phase
leads to un-quantized thermal Hall conductivities $ \kappa_{xy}/T $ even as $ T \rightarrow 0 $ limit.
However, as shown in the previous sections, the extra term vanishes at the Mirror symmetric point $ \beta=\pi/4 $ and also at the Abelian point,
so does the thermal Hall conductivity. So the thermal Hall conductivity should also change sign at the Mirror symmetric point.
Interestingly, as said above, the Heisenberg term $ J^y_H $ also changes from FM to AFM at $ \beta=\pi/4 $.
While the material is thought to be in the AFM side with $ \beta > \pi/4 $.
Both Z-x and Z-FM phases are gapped phases, to the quadratic order in Eq.\ref{Zxex} or Eq.\ref{gapmiddle}, thermal Hall conductivities  vanish.
However, expanding the dispersions to higher orders, various skewness \cite{thermalhigh} may move in,
even so, they show at most exponentially suppressed thermal Hall conductivities  $ \kappa_{xy}/T \sim c e^{-\Delta/T} $ where
the coefficient $ c $ is also suppressed at a low $ T $ and the $ \Delta $
is the gap in the two phases respectively.
These behaviours match those in the Zig-Zag and FM state in the $ \alpha-Ru Cl_3 $
in the low and high Zeeman field respectively.
It remains challeging for the $ (J,K,I) $ or the minimal $ (J,K, \Gamma ) $ model to naturally explain the Thermal Hall conductivities observed
in the experiments.

\section{Conclusions and discussions }

From symmetry analysis, plus some inputs from the  microscopic SWE calculations achieved in \cite{rh,rhh,rhtran},
we constructed various effective actions to describe the transitions
(1) The C-IC transition from the  Z-x to the IC-SkX at $ h_{c1} $:
    It has a single complex order parameter with the dynamic exponents $ z=2 $ and one Type-II dangerously irrelevant operator.
(2) The C-IC transition from  the Z-FM to the IC-SkX at $ h_{c2} $ in the middle of SOC $ \beta_1 < \beta < \beta_2 $:
   It has two complex order parameters with the dynamic exponents $ z=2 $ and two Type-II dangerously irrelevant operators.
(3) The C-C transition from  the Z-FM to the canted phase at $ h_{c2} $ in the left of SOC $ 0 < \beta < \beta_{1} $
     It takes a boosted form with the dynamic exponents $ z=1 $ and has an order parameter reduction (OPR) from two to one complex order parameter.
(4) Finally, the C-IC transition  from the canted to the IC-SkX at $ \beta_L $: It has an order parameter fractionization (OPF)
    from one to TWO complex order parameters with the dynamic exponent $ (z_x=3/2,z_y=3 ) $.
    (1) to (4) close the whole cycle in Fig.1.
All the 4 effective actions reach consistent descriptions on the IC-SkX phase centered in the phase diagram.
Our mean field analysis on these effective actions reproduced all the 5 ground states,
the quantum fluctuations above the mean field reproduced all the excitations such as Goldstone, roton and Higgs modes
above the ground states. Furthermore, we investigate the nature of all the 4 QPTs.

Recently, we also performed both microscopic and phenomenological effective actions to study Zeeman field induced
quantum phase transitions of spinor bosons in the presence of $ \pi $ flux \cite{pifluxgold,pifluxqsl}
or  in bosonic quantum Anomalous Hall systems \cite{NOFQD}.
Of course, the magnon condensations in the presence of SOC presented here with $ U(1)_{soc} $
and \cite{devil} without $ U(1)_{soc} $ is a different class of problems than the BEC in the presence of SOC in \cite{pifluxgold,pifluxqsl,NOFQD}.

 We identify carefully the relation between the quantum spin and the order parameters in various effective actions.
 It is important in the following way:

 (1) In the spinor boson case \cite{NOFQD}, the quantum spin is quadratically represented in terms of the two components
 complex order parameters $ \psi_1, \psi_2 $,
 it is the Zeeman field which directly tunes the relative magnitude between the two components.
 Here, near $ h=h_{c2} $, there are also  two components complex order parameters $ \psi_1, \psi_2 $,
 the Zeeman field $ h $ is implicitly embedded in the chemical potential term $ \mu=h_{c2}- h $.
 Counter-intuitively, as shown in Sec.IV, the $ \psi_1, \psi_2 $ always has equal amplitude, independent of the Zeeman field.

 (2) The ratio of the quantum spins in A/B sublattice is determined by the two generalized Bogliubov matrix elements $ u, v $.
 Even if getting to the $ \psi_{\pm} $ basis, it is always in the Ising limit where one of them vanishes, also independent of
 the Zeeman field. This is one of the crucial difference between the BEC of bosons and BEC of magnons:
 in the former, the quantum spin is quadratically represented in terms of the order parameter,
 in the latter, the quantum spin is linearly represented in terms of the order parameter, whose coefficients involve
 unitary transformation ( here below $ h_{c1} $ ) or Bogliubov transformation ( here above $ h_{c2} $ ).
 Even the two transformations are well defined only below $ h_{c1} $ or above $ h_{c2} $ respectively.
 The relation can be phenomenologically  continued into $ h_{c1} < h < h_{c2} $ and match the microscopic SWE calculations.

 (3) Although the universality classes of the transitions are completely determined by the order parameters,
 these relations are important to identify the correct spin-orbital orders of the states, also in  evaluating the spin-spin
 correlation functions inside all the 5 phases, also near the QCP at a finite temperature
 which can be directly detected by all kinds of Bragg spectroscopy \cite{lightatom1,braggbog,braggangle,braggeng,braggsingle,braggsoc}.

 (4) There is an order parameter fractionization (OPF) from one complex order parameter to two in the C-IC transition from canted
  to IC-SkX phase with the $ (z_x=3/2,z_y=3 ) $. There is also a dynamic exponent change from $ z=1 $ to $ z=2 $,
  Higgs mode to Roton mode. This fractionization is different, but related to the quantum spin fractionization into spinons plus a $ Z_2 $ flux
  when the systems gets into a quantum spin liquid phase from a magnetic ordered phase.
  For example, there could be a topological phase transition from a FM state to a $ Z_2 $ QSL  in the spinor bosons in the presence of $ \pi $ flux \cite{pifluxgold,pifluxqsl}.

 We also develop a new concept: Type-II dangerously irrelevant operators which considerably enrich the previously known
 Type-I dangerously irrelevant operators. It is instructive to look at the history associated with Type-II in different contexts:
 Type-II superconductors hosting a mixed vortex state in the presence of magnetic field was discovered after the Type-I superconductor.
 Type-II Weyl fermions \cite{weyl} hosting a Fermi-surface with non-zero density of states was discovered after the Type-I Weyl points with vanishing density of states.
 More recently, a Type-II deformation to a order from quantum disorder (OFQD) state leads to  the nearly OFQD (NOFQD) phenomena.
 While a Type-I deformation to a OFQD state acts trivially.

 One can summarize the two different sources of the extra "doppler" shift term \cite{doppler} in the Goldstone, roton or Higgs  modes in Eq.\ref{Goldhc1},\ref{GoldRoton}, \ref{GoldHiggs}.

 (1) For IC-momentum, it is due to the Type-II dangerously irrelevant operators, the number of which is equal to
 the number of IC- momenta condensations: one $ V $ near $ h_{c1} $  for Eq.\ref{Goldhc1} and two $ V_1, V_2 $ near $ h_{c2} $
 for Eq.\ref{GoldRoton}.

 (2) For C-momentum, it is due to boosted term in the kinetic energy. This is the case for Eq.\ref{GoldHiggs}.

 In fact, as shown in Sec.II and III, the effects of the dangerously irrelevant operators inside the symmetry broken
 IC-SkX phase  can be transformed into the boosted form inside the kinetic energy.

The spin-spin correlation functions in Z-x phase is evaluated in the appendix D. It can be similarly evaluated in all the other phases.
The qualitative behaviours of the thermal  Hall conductivity $ \kappa_{xy}/T $
in all the phases in Fig.1 or Fig.2 were outlined in VII. In view of its importance in the 4d 0r 5d Kitaev materials
\cite{halfinteger,unquantized}, the quantum Hall effects near $ \nu=5/2 $ \cite{NonFQHE} and underdoped cuprates \cite{cu,cu2}, following the
methods developed in \cite{ther1,ther2,ther3}, we will study its quantitative behaviours, especially in the QC regimes
in Fig.1 and Fig.2 in a separate publication \cite{S:un}

The $ U(1)_{soc} $ symmetry only holds along the $ (\alpha=\pi/2, \beta) $ SOC line and the longitudinal Zeeman field.
It may not hold in any general SOC systems. Here there could be many ways to break the $ U(1)_{soc} $ symmetry explicitly.
One way is to apply a transverse field as discussed in appendix E.
Another way is to look at a generic  $ (\alpha,\beta) $, or one can apply both at the same time \cite{partial}.
In \cite{devil}, we studied various magnon condensations  in a generic $ (\alpha,\beta) $ which has no $ U(1)_{soc} $ symmetry.
As expected, it is quite different than the magnon condensation with the $ U(1)_{soc} $ symmetry addressed in this paper.
Some crucial differences between the two were spelled out in the appendix F in \cite{devil}.
Especially, we showed that the Z-x state remains stable in a large SOC parameter regime near $ \pi/2 $, just changes from the exact to the classical ground state. In fact, it is the most robust
quantum phase in the whole global phase diagram in the generic $ (\alpha,\beta) $.  So we expect some features
in Fig.1 will remain when the $ U(1)_{soc} $ symmetry is broken.
It would be interesting to look at how the phases and QPTs in Fig.1, especially the IC-SkX phase evolve
when the $ U(1)_{soc} $ symmetry was explicitly broken.
Of course, the Goldstone mode inside the canted phase and the IC-SkX phase will be gapped due to
the explicit $ U(1)_{soc} $ symmetry breaking \cite{angleH}.

As mentioned at the end of Sec.VII, if extending the results to a honeycomb lattice with either bosons or fermions,
the IC-SkX likely melts into a QSL, then the transitions at $ h_{c1} $ and $ h_{c2} $ will become
two Topological transitions driven by the condensation of spinons or flux or some sort of fermions.
Then it may be directly relevant to current trends searching for QSL driven by a Zeeman field in 4d or 5d Kitaev materials.

{\bf Acknowledgements}

We thank Dapeng Yu for hospitality during the authors visit at Institute for Quantum Science and Engineering, Shenzhen 518055, China.
J.Ye thank Dr. Zhong Ruidan for experimental data related discussions.

\appendix

\section{ The relations of the symmetry operators between the previous works and the present }

 The  RH model  with generic $0<\beta<\alpha<\pi/2$  in the original basis is:
\begin{align}
	\mathcal{H}=
	-J\sum_i[S_iR_x(2\alpha)S_{i+x}
		+S_iR_y(2\beta)S_{i+y}]
\end{align}

   Its symmetry are listed as \cite{rh,rhh,rhtran,devil}
\begin{enumerate}
	\item The Time reversal $\mathcal{T}: S\to -S$
		and $k\to -k$ ( or equivalently $(x,y)\to(x,y)$ ).
    \item The translational symmetry:	
	\item The  three spin-orbital coupled Z$_2$ symmetries \cite{footnote}:
	\begin{enumerate}
		\item $\tilde{\mathcal{P}}_x:
		(S^x,S^y,S^z)\to(S^x,-S^y,-S^z)$
		and $(k_x,k_y)\to(k_x,-k_y)$;
		\item $\tilde{\mathcal{P}}_y:
		(S^x,S^y,S^z)\to(-S^x,S^y,-S^z)$
		and $(k_x,k_y)\to(-k_x,k_y)$;
		\item $\tilde{\mathcal{P}}_z:
		(S^x,S^y,S^z)\to(-S^x,-S^y,S^z)$
		and $(k_x,k_y)\to(-k_x,-k_y)$;
	\end{enumerate}		
\end{enumerate}

   The RH model with $ ( \alpha=\pi/2, \beta) $ in a Zeeman field ( in the $R_x(\pi/2)$-rotated basis) is:
\begin{align}
	\mathcal{H}=
	-J\sum_i[S_iR_x(\pi)S_{i+x}
		+S_iR_z(2\beta)S_{i+y}]
		-H\sum_iS_i^z
\end{align}
  whose symmetries are classified in \cite{rh,rhh,rhtran} as:
\begin{enumerate}
	\item  The translational symmetry:
	\item  The spin-orbital $ U(1)_{soc} $ symmetry:
	\item  The three spin-orbital coupled Z$_2$ symmetries:
	\begin{enumerate}
		\item $\mathcal{T}\circ\tilde{\mathcal{P}}_x:
		(S^x,S^y,S^z)\to(-S^x,S^y,S^z)$
		and $(k_x,k_y)\to(-k_x,k_y)$ ( equivalently $(x,y)\to(x,-y)$ ).		
		\item $\mathcal{T}\circ\tilde{\mathcal{P}}_y:
		(S^x,S^y,S^z)\to(S^x,-S^y,S^z)$
		and $(k_x,k_y)\to(k_x,k_y)$; ( equivalently  $(x,y)\to(-x,-y)$ ).
		\item $\tilde{\mathcal{P}}_z:
		(S^x,S^y,S^z)\to(-S^x,-S^y,S^z)$
		and $(k_x,k_y)\to(-k_x,k_y)$;
	\end{enumerate}	
	\item The space reflection with respect to the y axis:
		$\mathcal{I}_y: S\to S$ and $(x,y)\to (-x,y)$ ( equivalently  $(k_x,k_y)\to (-k_x,k_y)$ ) which is
    the enlarged symmetry at $ \alpha=\pi/2 $ absent at the generic $ (\alpha, \beta) $ discussed above.
    This  enlarged symmetry at $ \alpha=\pi/2 $ was  missed  in \cite{rh,rhh,rhtran}.
\end{enumerate}
It is easy to see the following relations:
$\mathcal{T}\circ\tilde{\mathcal{P}}_x
=\mathcal{T}\circ\mathcal{I}_x\circ\mathcal{P}_x$,
$\mathcal{T}\circ\tilde{\mathcal{P}}_y
=\mathcal{T}\circ\mathcal{I}_x\circ\mathcal{I}_y\circ\mathcal{P}_y$, and
$\mathcal{I}_y\circ\tilde{\mathcal{P}}_z
=\mathcal{P}_z$. In view of $\mathcal{I}_y $ is conserved,  the three spin-orbital coupled Z$_2$ symmetries
can be simplified as
$ \mathcal{T}\circ\mathcal{I}_x\circ\mathcal{P}_x, \mathcal{T}\circ\mathcal{I}_x \circ\mathcal{P}_y, \mathcal{P}_z $
which are identical to those listed in Sec.I.

\section{Spin-wave expansion to order $1/S $.}

  We first review some results from spin-wave expansion (SWE) performed in \cite{rhh}.
  Especially, we stress the unitary transformation in the Z-x state below $ h_{c1} $ and the FM state above $ h_{c2} $
  which are crucial to derive the relations between the quantum spin and the order parameters inside
  the IC-SkX phase. As presented in Sec.II and III, the former is from bottom-up and the latter is from top-down.
\subsection{ Unitary transformation in the Z-x state in low field}

 In a weak magnetic field $ h < h_{c1} $, the ground-state is the Z-x state with the classical spin configuration:
\begin{align}
	\mathbf{S}_i=S(0,0,(-1)^{i_x})
\end{align}
Performing the Holstein Primakoff transformation
for $A/B$ sublattice \cite{rhh}
\begin{align}
	&S_i^+=\sqrt{2S}\Big(1-\frac{1}{2}\frac{n_i}{2S}+\cdots\Big)a_i,\quad
	S_i^-=\sqrt{2S}a_i^\dagger\Big(1-\frac{1}{2}\frac{n_i}{2S}+\cdots\Big),\quad
	S_i^z=S-a_i^\dagger a_i,\quad i\in A;   \\  \nonumber
	&S_j^+=\sqrt{2S}b_j^\dagger\Big(1-\frac{1}{2}\frac{n_j}{2S}+\cdots\Big),\quad
	S_j^-=\sqrt{2S}\Big(1-\frac{1}{2}\frac{n_j}{2S}+\cdots\Big)b_j,\quad
	S_j^z=-S+b_j^\dagger b_j,\quad j\in B.
\end{align}
In momentum space, the Hamiltonian takes the form
\begin{align}
	\mathcal{H}=&-2NJs^2
	+H\sum_k(a_k^\dagger a_k-b_k^\dagger b_k)
	+4JS\sum_k(a_k^\dagger a_k+b_k^\dagger b_k)  \nonumber  \\
	&-2JS\sum_k[\cos k_x a_k^\dagger b_k+\cos k_x b_k^\dagger a_k
	+\cos(k_y-2\beta)a_k^\dagger a_k
	+\cos(k_y+2\beta)b_k^\dagger b_k]
\end{align}
 By performing a unitary transformation
\begin{align}
	a_\mathbf{k}=s_\mathbf{k}\alpha_\mathbf{k}
	+c_\mathbf{k}\beta_\mathbf{k},\quad
	b_\mathbf{k}=s_\mathbf{k}\beta_\mathbf{k}
	-c_\mathbf{k}\alpha_\mathbf{k},
\label{unitaryTrans}
\end{align}
where $s_\mathbf{k}=\sin(\theta_{k,h}/2)$, $c_\mathbf{k}=\cos(\theta_{k,h}/2)$,
and $\tan\theta_{k,h}=\cos k_x/(\sin2\beta\sin k_y-h)$, the Hamiltonian can be put in the diagonal form:
\begin{align}
	\mathcal{H}=-2NJS^2
	+4JS\sum_k[\omega_{+}(k)\alpha_k^\dagger\alpha_k+\omega_{-}(k)\beta_k^\dagger\beta_k]
\label{betak}
\end{align}
where $ k $ is in the reduced BZ and the excitation spectrum is:
\begin{align}
	\omega_{\pm}(k)=1-\frac{1}{2}\cos2\beta\cos k_y
	\pm\frac{1}{2}\sqrt{\cos^2 k_x+(\sin 2\beta\sin k_y-h)^2}.
\end{align}

  The unitary transformation matrix elements $s_\mathbf{k}=\sin(\theta_{k,h}/2)$ and $c_\mathbf{k}=\cos(\theta_{k,h}/2)$
  are useful to establish the connections between the transverse quantum spin and the order parameter near $ h_{c1} $ in
  Eq.\ref{ordercs}.

\subsection{ Bogoliubov transformation in the FM in the high field}
In a strong magnetic field, the ground-state is Z-FM state with the classical spin configuration:
\begin{align}
	\mathbf{S}_i=S(0,0,1)
\end{align}
Performing the standard Holstein Primakoff transformation
\begin{align}
	&S_i^+=\sqrt{2S}\Big(1-\frac{1}{2}\frac{n_i}{2S}+\cdots\Big)a_i,\quad
	S_i^-=\sqrt{2S}a_i^\dagger\Big(1-\frac{1}{2}\frac{n_i}{2S}+\cdots\Big),\quad
	S_i^z=S-a_i^\dagger a_i.
\end{align}
In momentum space, the Hamiltonian takes the form
\begin{align}
	\mathcal{H}=-NHS+H\sum_{k}a_k^\dagger a_k
	-JS\sum_k[2\cos (k_y-2\beta)a_k^\dagger a_{k}
	+\cos k_x(a_ka_{-k}+a_k^\dagger a_{-k}^\dagger)]
\end{align}
 By introducing the Bogoliubov transformation as
\begin{align}
	a_\mathbf{k}=u_\mathbf{k}\alpha_\mathbf{k}
	+v_\mathbf{k}\alpha_{-\mathbf{k}}^\dagger,\quad
	a_{-\mathbf{k}}^\dagger=v_\mathbf{k}\alpha_{\mathbf{k}}
	+u_\mathbf{k}\alpha_{-\mathbf{k}}^\dagger.
\label{bogoliubovTrans}
\end{align}
where $u_\mathbf{k}=\cosh\eta_k$, $v_\mathbf{k}=\sinh\eta_k$ and
$\tanh2\eta_k=\cos k_x/(h-\cos2\beta\cos k_y)$,
the Hamiltonian takes the diagonal form
\begin{eqnarray}
	\mathcal{H}&=&-NH(S+\frac{1}{2})+JS\sum_k
	[\omega(k)\alpha_k^\dagger \alpha_k
	+\omega(-k)\alpha_{-k}\alpha_{-k}^\dagger]   \\ \nonumber
	&=&-NH(S+\frac{1}{2})+JS\sum_k \omega_k
	+2JS\sum_k \omega_k\alpha_k^\dagger \alpha_k
\label{alphak}
\end{eqnarray}
where $ k $ is in the BZ and the spin wave dispersion is
\begin{equation}
	\omega_k=\sqrt{(h-\cos2\beta\cos k_y)^2-\cos^2 k_x}-\sin2\beta\sin k_y
\label{eq:Ek}
\end{equation}

 The Bogoliubov transformation matrix elements $u_\mathbf{k}$ and $v_\mathbf{k} $
 are useful to establish the connections between the transverse quantum spin and the order parameter near $ h_{c2} $ in Eq.\ref{orderuv}.

\section{ Microscopic SWE calculations on the A/B sublattice  ratio near $ h_{c1}, h_{c2} $ and $ h_L $. }

In the classic limit ($S\to\infty$),
one can take the general ansatz of the IC-SkX state:
\begin{align}
	&S_i=
	S(\sin\theta_A\cos(\phi_A-k_0i_y),
	  \sin\theta_A\sin(\phi_A-k_0i_y),
	  \cos\theta_A),\quad i\in A ~~(i_x \text{ is odd}) \\ \nonumber
	&S_j=
	S(\sin\theta_B\cos(\phi_B+k_0j_y),
	  \sin\theta_B\sin(\phi_B+k_0j_y),
	  \cos\theta_B),\quad j\in B ~~(j_x \text{ is even})
\end{align}
   which is equivalent to:
\begin{align}
	S_i^z&=(S/2)[\cos\theta_A+\cos\theta_B+(-1)^{i_x}(\cos\theta_A-\cos\theta_B)]   \\  \nonumber
	S_i^+&=(S/2)[\sin\theta_A+\sin\theta_B+(-1)^{i_x}(\sin\theta_A-\sin\theta_B)]e^{(-1)^{i_x}i(\phi_0+k_0i_y)}
\label{icskx}
\end{align}

Then the ground-state energy is
\begin{align}
    E_{GS}=\min_{\theta_A,\theta_B,k_0}
	    NJS^2[\cos(\theta_A+\theta_B)
	    +\sin^2(\beta+k_0/2)\sin^2\theta_A
	    +\sin^2(\beta-k_0/2)\sin^2\theta_B
	    -h(\cos\theta_A+\cos\theta_B)	
	    -1]
\end{align}
The minimization procedure automatically
gives  $\theta_A,\theta_B,k_0$ for the IC-SkX state which reduces to the
FM state, canted state and Z-x state in the corresponding $(\beta,h)$ regime.

Near $ h_{c2} $, it gives $\lim_{h\to h_{c2}^-}\theta_A=\lim_{h\to h_{c2}^-}\theta_B=0$, so
Eq.\ref{icskx} reduces to the Z-FM state with the non-trivial ratio:
\begin{align}
	\lim_{h\to h_{c2}^-}\frac{\sin\theta_A}{\sin\theta_B}
	=\lim_{h\to h_{c2}^-}\frac{h-\sqrt{3h^2-1-h^4}}{h^2-1}
	=\sqrt{\sin^42\beta+\sin^22\beta}
	-\sqrt{\sin^42\beta-\cos^22\beta}
\label{hc2ratio}
\end{align}
  which matches Eq.\ref{hc2ratio} achieved by the effective action from $ h^{+}_{c2} $.

thus it is easy to verify
\begin{align}
    \lim_{h\to h_{c2}^-}\frac{\sin\theta_A}{\sin\theta_B}
	\Big|_{\beta=\beta_1}=1,
	\quad
    \lim_{h\to h_{c2}^-}\frac{\sin\theta_A}{\sin\theta_B}
	\Big|_{\beta=\pi/4}=\sqrt{2}-1,	
\end{align}

Near $ h_{c1} $, it gives $\lim_{h\to h_{c1}^+}\theta_A=0$,
$\lim_{h\to h_{c1}^+}\theta_B=\pi$, so
Eq.\ref{icskx} reduces to the Z-x state with the non-trivial ratio:
\begin{align}
	\lim_{h\to h_{c1}^+}\frac{\sin\theta_A}{\sin\theta_B}
	=\lim_{h\to h_{c1}^+}
	[2-\cos2\beta\cos k_0-\sqrt{(2-\cos2\beta\cos k_0)^2-1}]
\label{hc1ratio}
\end{align}
which matches Eq.\ref{hc1ratiocs} achieved by the effective action from $ h^{-}_{c1} $.

thus it is easy to verify
\begin{align}
    \lim_{h\to h_{c1}^+}\frac{\sin\theta_A}{\sin\theta_B}
	\Big|_{\beta=0}=1,
	\quad
    \lim_{h\to h_{c1}^+}\frac{\sin\theta_A}{\sin\theta_B}
	\Big|_{\beta=\pi/4}=2-\sqrt{3},	
\end{align}

 The Ic-momentum along $h_{c2}$ is found to be:
\begin{align}
    k_0=\arccos[\cot2\beta\sqrt{1+\sin^22\beta}]\sim [40(\sqrt{5}-1)]^{1/4}\sqrt{\beta-\beta_1}
\end{align}
  where the second equation works near $ \beta_1 $.

  Near $\beta_L$, Eq.\ref{icskx} reduces to the canted state with the non-trivial ratio:
\begin{align}
	\frac{\sin\theta_A}{\sin\theta_B}
	=1-\frac{\sin k_0}{\tan2\beta_L}+O(k_0^2)
\end{align}

At $h=1$, $\cos2\beta_L$ is root of equation $z^4+z^3+2z-2=0$.

\section{ The spin-spin correlation functions in the Z-x state below $ h_{c1} $ }

  Using the relations between the quantum spin and the order parameters, one can evaluate
  the spin-spin correlations functions in all the phases from the effective actions, especially their scaling functions near the QCPs.
  Here, for simplicity, we just evaluate them it in the Z-x state.

   The low-energy effective action corresponding to the Hamiltonian Eq.\ref{betak} can be written as
\begin{align}
        S_\text{eff}=\int_0^\beta d\tau \sum_k[\beta_{k,\tau}^*\partial_\tau\beta_{k,\tau}+4JS\omega_{h,k}^-\beta_{k,\tau}^*\beta_{k,\tau}]
        =\sum_{k,\omega_n}[-i\omega_n+4JS\omega_{h,k}^-]\beta^*(k,i\omega_n)\beta(k,i\omega_n)
\end{align}
which leads to the only non-vanishing correlation function:
\begin{align}
      \chi_{\beta\beta^*}(k,i\omega_n)=   \langle \beta(k,i\omega_n)\beta^*(k,i\omega_n)\rangle=\frac{1}{-i\omega_n+4JS\omega_{h,k}^-}
\label{chibeta}
\end{align}

  The relation between the quantum spin and the order parameter Eq.\ref{k0iyjy} (or Eq.\ref{ordercs} ) leads to
\begin{align}
        \langle S_A^+(k,i\omega_n)S_A^-(k,i\omega_n)\rangle=
        \langle S_B^-(k,i\omega_n)S_B^+(k,i\omega_n)\rangle
        =S(1+\cos\theta_0)\langle \beta(k,i\omega_n)\beta^*(k,i\omega_n)\rangle
\end{align}
and
\begin{align}
        \langle S_B^-(k,i\omega_n)S_A^-(k,i\omega_n)\rangle
        =\langle S_A^+(k,i\omega_n)S_B^+(k,i\omega_n)\rangle
        =S\sin\theta_0\langle \beta(k,i\omega_n)\beta^*(k,i\omega_n)\rangle
\end{align}
  where  $\theta_{h,k}=\theta_{h,k=Q}=\theta_0$  are evaluated at the minimal $\mathbf{K}_0=(0,k_0)$.

  The  analytical continuation of Eq.\ref{chibeta} leads to
\begin{align}
        \mathrm{Im}[\chi_{\beta\beta^*}(k,i\omega_n\to\omega+i0^+)]=\pi\delta(4JS\omega_{h,k}^--\omega)
\end{align}
 which leads to the equal-time correlation function by the fluctuation-dissipation theorem:
\begin{align}
        S_{\beta\beta^*}(k)=\int\frac{d\omega}{2\pi}\frac{-2\mathrm{Im}[\chi_{\beta\beta^*}(k,\omega)]}{1-e^{-\omega/T}}
        =1-\frac{1}{e^{4JS\omega_{h,k}^-/T}-1}
\end{align}
 which leads to the equal-time SSCFs (  or structure factor ):
\begin{align}
        S_{AA}^{+-}(k)=S_{BB}^{-+}(k)=S(1+\cos\theta_0)\Big[1-\frac{1}{e^{4JS\omega_{h,k}^-/T}-1}\Big]   \\  \nonumber
        S_{AB}^{++}(k)=S_{BA}^{--}(k)=S\sin\theta_0\Big[1-\frac{1}{e^{4JS\omega_{h,k}^-/T}-1}\Big]
\end{align}

 Following \cite{rh}, one can define the uniform spin $M=(S_A+S_B)/2$  and the staggered spin $M=S_A-S_B$,
then
\begin{align}
        S_u^{+-}(k)=\frac{1}{4}S_{AA}^{+-}(k)
                            =\frac{1}{2}S\cos^2(\theta_0/2)\Big[1-\frac{1}{e^{4JS\omega_{h,k}^-/T}-1}\Big]   \\  \nonumber
        S_u^{++}(k)=-S_{AB}^{++}(k)
                            =-S\sin\theta_0\Big[1-\frac{1}{e^{4JS\omega_{h,k}^-/T}-1}\Big]
\end{align}
which, after exchanging the order of the spin operators, match those achieved in \cite{rh} by the spin wave expansion.

For a generic $ ( \alpha,\beta) $ in Eq.1 which breaks the $ U(1)_{soc} $ explicitly, identifying the low energy modes is much more involved,
so evaluating the spin-spin correlation functions is much more involved in \cite{devil}.

\section{ Order parameter and the QPT  in the transverse fields}

As mentioned in the introduction, due to the SOC, the response to a Zeeman field depends on the orientation of the field.
Here, we apply a transverse field \cite{rhtran} to the RFHM in Eq.M1:
\begin{align}
    \mathcal{H}
	=\mathcal{H}_\text{RH}-\mathbf{H}\cdot\sum_i\mathbf{S}_i\>.
\end{align}
where the two transverse fields are
$\mathbf{H}\parallel \hat{x}$ or $\mathbf{H}\parallel \hat{y}$. Both break the $ U(1)_{soc} $.
So it should be quite different than the case with $ U(1)_{soc} $. Indeed,
as shown in \cite{rhtran}, in contrast to the longitudinal case \cite{rhh}, the C- magnons always emerge out
in the competition against the IC- magnons and drive the QPT.
The magnon condensation  by the SWE in \cite{rhtran} suggests the order parameter takes the form:
\begin{equation}
 \langle\alpha_\mathbf{k}\rangle=\psi \delta_{\mathbf{k},\mathbf{Q}}
\end{equation}
where $\mathbf{Q}=(\pi,0)$ is the C- condensation momentum.

  This type of magnon condensation leads to the transverse spin components:
\begin{align}
	\langle S_i^z+iS_i^y\rangle
	\propto (-1)^{i_x}(u\psi +v\psi^\ast)
	\propto (-1)^{i_x}(\psi+\psi^\ast),
	\text{ for } \mathbf{H}\parallel \hat{x}  \\ \nonumber
	\langle S_i^z+iS_i^x\rangle
	\propto (-1)^{i_x}(u\psi+v\psi^\ast)
	\propto (-1)^{i_x}(\psi+\psi^\ast),
	\text{ for } \mathbf{H}\parallel \hat{y}
\label{rhtranreal}
\end{align}
where, as alerted in Sec.IV, $u=u_{\mathbf{Q}}=\infty$ and $v=v_{\mathbf{Q}}=\infty$,
but $u/v=1$, so can be simply factored out in the relation between the quantum spin and the order parameter.
The above equation suggests the order parameter can be taken as one REAL field $\phi=\psi+\psi^\ast$.

The symmetry of the Hamiltonian $\mathcal{H}$
in $H_x$ is generated by
1) translation $\mathcal{T}_x$ and $\mathcal{T}_y$;
2) space reflection $\mathcal{I}_y$;
3) spin-orbital reflection
$\mathcal{I}_x\circ\mathcal{P}_x$,
$\mathcal{T}\circ\mathcal{I}_x\circ\mathcal{P}_y$,
$\mathcal{T}\circ\mathcal{P}_z$.
The translation takes $\phi(x,y)\to\phi(x,y)$,
and space reflection $\mathcal{I}_y$ takes $\phi(x,y)\to\phi(-x,y)$,
but spin-orbital reflection
$\mathcal{I}_x\circ\mathcal{P}_x:\phi(x,y)\to-\phi(x,-y)$,
$\mathcal{T}\circ\mathcal{I}_x\circ\mathcal{P}_y:\phi(x,y)\to\phi(x,-y)$,
$\mathcal{T}\circ\mathcal{P}_z:\phi(x,y)\to-\phi(x,y)$.

The symmetry of the Hamiltonian $\mathcal{H}$
in $H_y$  has the same 1) and 2),
but 3) spin-orbital reflections become
$\mathcal{I}_x\circ\mathcal{P}_y$,
$\mathcal{T}\circ\mathcal{I}_x\circ\mathcal{P}_x$,
$\mathcal{T}\circ\mathcal{P}_z$.
The translation takes $\phi(x,y)\to\phi(x,y)$,
and space reflection $\mathcal{I}_y$ takes $\phi(x,y)\to\phi(-x,y)$,
but spin-orbital reflection
$\mathcal{I}_x\circ\mathcal{P}_y:\phi(x,y)\to-\phi(x,-y)$,
$\mathcal{T}\circ\mathcal{I}_x\circ\mathcal{P}_x:\phi(x,y)\to\phi(x,-y)$,
$\mathcal{T}\circ\mathcal{P}_z:\phi(x,y)\to-\phi(x,y)$.

Obviously, the high field X-FM or Y-FM state breaks no symmetries of the corresponding Hamiltonian.
So the above symmetry analysis leads to the effective action with $ z=1 $:
\begin{align}
    \mathcal{S}
    =\int d\tau d^2r [
	(\partial_\tau\phi)^2
	+v_x^2(\partial_x\phi)^2
	+v_y^2(\partial_y\phi)^2
	-\mu\phi^2+u\phi^4]
\end{align}
where $\phi $ is a real scalar field and the SWE in \cite{rhtran} shows $ \mu= h_{c}-h $.
Note that  the spin-orbital $ Z_2 $  reflection,
i.e. $\mathcal{T}\circ\mathcal{P}_z:\phi\to-\phi$ dictates the absence of odd power of $\phi $ terms.
This is nothing but standard 3D Ising universality class.
In contrast to the longitudinal Zeeman field case, there is no intermediate phases, so just one transition.

  At mean field level, $\psi=m$
\begin{align}
	\mathcal{S}_0=-\mu m^2+um^4
\end{align}
When $\mu=h_{c}-h  <0$, $m=0$, which means the fully polarized spin state
$\langle \mathbf{S}_i\rangle=S(1,0,0)$ for $\mathbf{H}\parallel \hat{x}$
and $\langle \mathbf{S}_i\rangle=S(0,1,0)$ for $\mathbf{H}\parallel \hat{y}$.
When $\mu>0$, $m^2=\mu/2u$, which means the canted phase
$\langle \mathbf{S}_i\rangle=(\sqrt{S^2-m^2},0,(-1)^{i_x}m)$
for $\mathbf{H}\parallel \hat{x}$ and
$\langle \mathbf{S}_i\rangle=(0,\sqrt{S^2-m^2},(-1)^{i_x}m)$
for $\mathbf{H}\parallel \hat{y}$.

When $\mu<0$, the action to the quadratic order:
\begin{align}
    \mathcal{S}_2
	=\int d\tau d^2r[
	(\partial_\tau\phi)^2
	+v_x^2(\partial_x\phi)^2
	+v_y^2(\partial_y\phi)^2
	-\mu\phi^2]
	\Longrightarrow
	\omega_k=\sqrt{-\mu+v_x^2k_x^2+v_y^2k_y^2}
\end{align}
 which coincides with the spectrum inside the FM phase achieved by SWE in \cite{rhtran}.

When $\mu>0$, expanding $\phi=m+\delta\phi$ to the quadratic order
\begin{align}
    \mathcal{S}_2
	=\int d\tau d^2r[
	(\partial_\tau \delta \phi)^2
	+v_x^2(\partial_x  \delta \phi  )^2
	+v_y^2(\partial_y  \delta \phi   )^2
	+(6m^2u-\mu)  \delta \phi^2]
	\Longrightarrow
	\omega_k=\sqrt{(6m^2u-\mu)+v_x^2k_x^2+v_y^2k_y^2}
\end{align}
which may also be rewritten as
$\omega_k=\sqrt{2\mu+v_x^2k_x^2+v_y^2k_y^2}$. It coincides with the spectrum inside the canted phase achieved by SWE in \cite{rhtran}.

It is easy to show that the QPTs at the special Abelian points in the two transverse field cases is similar as that in the
longitudinal case studied in Sec.IV, so it is in the 3D XY universality class \cite{exclude2}. Again, despite there are two C-
momenta condensation, the order parameter reduction (OPR)
mechanism discovered in Sec.IV applies here: there is only one complex order parameter at
the Abelian point. However, any small SOC breaks the $ U(1) $ symmetry at the Abelian point to a $ Z_2 $ symmetry, open a gap to the imaginary part and
picks up the REAL component in Eq.\ref{rhtranreal} as the critical mode, therefore transfers the 3D XY class to the 3D Ising class.
In a sharp contrast, in the longitudinal case, any small SOC in Fig.1 still keeps the $ U(1)_{soc} $ symmetry at the Abelian point,
just breaks the $ Z_2 $ symmetry listed below Eq.\ref{boostleft}, therefore generates a boost to the 3d XY model in the $ \partial_{\tau} $ term
as shown in Sec.IV-C.

The generic $ (\alpha,\beta) $ case also breaks  the $ U(1)_{soc} $ symmetry explicitly and
the order parameter is also real \cite{devil}. However, in contrast to the transverse field case where only C-magnon condensations can happen,
there are both C- and IC-magnons condensations which lead to different QPTs and different spin-ordered phases shown in \cite{devil}.

\section{ Order parameter and the QPT of the AFM in a uniform field}

For an AFM in a uniform field,
a big $h$ leads to a fully polarized state,
Z-FM state, which is not only the ground state but also an exact eigenstate.
A Simple spin-wave calculation shows $\omega\sim \Delta+ v^2 k^2$ near $(\pi,\pi)$.
In this case, because it is an exact eigenstate, neither Bogoliubov transformation nor unitary transformation is needed,
thus the relation between the spin  and the order parameter is simply
$\langle S_i^+\rangle= (-1)^{i_x+i_y}\psi$ with a complex field $\psi$.
The effective action consistent with the $ U(1)_s $ symmetry has $ z=2 $:
\begin{align}
	\mathcal{S}=\int d\tau d^2r
	[\psi^*\partial_\tau\psi+v^2|\nabla\psi|^2-\mu|\psi|^2+U|\psi|^4]
\end{align}
which belongs to $z=2$ zero density SF-Mott transition universality class, therefore confirm the assumption used in \cite{z2}.

When $\mu<0$, $\psi=0$ the mean field ground state is Z-FM state.
While $\mu>0$, $\psi=me^{i\phi_0}$ the mean field ground state is canted state
$$\mathbf{S}_i
=((-1)^{i_x+i_y}m\cos\phi_0,
(-1)^{i_x+i_y}m\sin\phi_0,
\sqrt{S^2-m})$$
 which supports one gapless Goldstone mode due to the $ U(1)_s $ symmetry breaking.
 It leads to a un-quantized thermal conductivity $ \kappa_{xx}/T $ even at the $ T \rightarrow 0 $ limit, but
 thermal Hall conductivity vanishes.

As mentioned in Sec.IV-C and also appendix E, at the Abelian point $\beta=0$, $ c=0 $ in Eq.\ref{boostleft},
the Hamiltonian has an additional space reflection with respect to $x$ axis,
thus $c=0$  belongs to standard 3D XY universality class with the dynamic exponent $ z=1 $ which
is dramatically different than the above case.


\begin{thebibliography}{99}
\bibitem{symmetrydiffer} Here  when classifying the symmetries, we use a slightly different notation  than that used in the previous works \cite{rh,rhh,rhtran,rafhm}. The only symmetry not realized is the space reflection with respect to $ y $ axis $ {\cal I}_y $.
    See appendix A for details.
\bibitem{iclead} The in-commensurate is always important in SOC systems, for example, it leads to the gapless 1d Luttinger liquids
embedded in 2d in \cite{devil}. It also leads to dramatic effects in spinor bosons condensing at in-commensurate momenta \cite{SFnon,SFQAH}.
\bibitem{morederivative} Note that our microscopic calculations in \cite{rhh} also find more higher derivative terms
such as  $ v_{xx} (\partial_x^2\phi)^2 + v_{yy} (\partial_y^2\phi)^2 +  v_{xy} \partial_x^2 \phi \partial_y^2\phi $, but
including these extra terms will not change the results derived in the following, so we will not write them out explicitly.
\bibitem{contrastNOFQD} The effective action Eq.\ref{pmbasisM} can be contrasted with that in \cite{NOFQD} to describe the
quantum phase transition induced by a Zeeman field due to the nearly order from quantum disorder (NOFQD) phenomenon.
\bibitem{irrational} When $k_0/\pi$ is an irrational number, then $ e^{i k_0 n } $ takes all the possible numbers
 with modular 1 when $ n=0, 1,2,3.......$.
 However, when $k_0/\pi$ is a rational number such as $ 1/2 $, then $  e^{i k_0 n } $ can only take 4 values $ 1, i,-1,-i $.
 When drawing analogy to classical chaos, the former maps to chaotic orbits, the latter to periodic ones.
\bibitem{exclude} In fact, one may intend to add a $ B (|\psi_1|^2-|\psi_2|^2)^2 $ term. However,
 under $ U(1)\textsubscript{soc} $, it changes as $
	(|\psi_1|^2-|\psi_2|^2)^2
	=(\psi_+\psi_-^\ast+\psi_+^\ast\psi_-)^2
\to
	(e^{+2i\phi_0}\psi_+\psi_-^\ast+e^{-2i\phi_0}\psi_+^\ast\psi_-)^2
	=\cos^2(2\phi_0)(\psi_+\psi_-^\ast+\psi_+^\ast\psi_-)^2
	+\sin^2(2\phi_0)(\psi_+\psi_-^\ast-\psi_+^\ast\psi_-)^2 $.
     So this term is absent in the effective action.
\bibitem{notOPF} Here when we say OPF, it means the order parameter splits from one to two in the C-IC transition from the canted phase to the IC-SkX phase. In the C-IC transition from the Z-x phase to the IC-SkX at $ h_{c1} $, there is always one complex order parameter.
    In the C-IC transition from the Z-FM phase to the IC-SkX at $ h_{c2} $, there are always two complex order parameters.
    So there is no OPF in the latter two cases.
\bibitem{footnote} The original $\mathcal{P}$ in \cite{rh,rhh,rhtran,devil} is denoted by $\tilde{\mathcal{P}}$ here.
\bibitem{exclude2} Excluding the trivial Abelian point at $ \beta=0 $ in the $ H_x $ case which is nothing but a FM in a uniform field.
\bibitem{admit} Its solution can be written as $ \psi_1+\psi_1^\ast= Re [\tilde{\psi} e^{i(\phi_0+k_0y)}],
\psi_2-\psi_2^\ast= i Im [\tilde{\psi} e^{i(\phi_0+k_0y)} ]$. They lead to $ \psi_1= \tilde{\psi} e^{i(\phi_0+k_0y)} + i \chi_2 $
and $ \psi_2= \tilde{\psi} e^{i(\phi_0+k_0y)} +  \chi_1 $ where $ \chi_1, \chi_2 $ are any two real functions.
Thus $ \psi_2 -\psi_1= \chi_1-i \chi_2 =\chi $ is any complex function. It leads to  the physical solution
$\psi_\alpha=\tilde{\psi}_\alpha e^{ik_0y}, \alpha=1,2 $ where $ \tilde{\psi}_1, \tilde{\psi}_2
= \tilde{\psi}_1 + \chi e^{-ik_0y} $ are two independent complex functions.
They are nothing but the two complex order paremeters inside the IC-SkX phase.
\bibitem{angleH} In the experiments \cite{halfinteger,unquantized}, there is very little anisotropy in the direction of the
in-plane magnetic field, but dramatic difference between the in-plane and c-axis. J.Ye thank Dr. Zhong Ruidan for alerting us this fact.
Our case is extremely sensitive to the in-plane angle also due to the special $ U(1)_{soc} $.
However, after breaking the  $ U(1)_{soc} $ symmetry explicitly, the sensitivity  drops, but still depeends on the in-plane angle.
\end{thebibliography}
\end{document}